\documentclass[review,3p,12pt]{elsarticle}

\usepackage{graphicx}
\usepackage{color}
\usepackage{epsfig}
\usepackage{booktabs}
\usepackage{latexsym}
\usepackage{amsmath}
\usepackage{amssymb}
\usepackage{mathtools}
\usepackage{hyperref}
\usepackage{subfigure}
\usepackage{ulem}
\usepackage{lettrine}
\usepackage{verbatim}
\usepackage{array}
\usepackage{varwidth}
\usepackage{mathptmx} 
\usepackage{setspace}

\biboptions{numbers,compress}

\journal{Computer \& Mathematics with applications}

\begin{document}

\begin{frontmatter}

\title{A Lattice Boltzmann dynamic-Immersed Boundary scheme for the transport of deformable inertial capsules in low-Re flows}

\author[unibas]{Alessandro Coclite\corref{cor}}
\ead{alessandro.coclite@unibas.it}

\author[dimeg]{Sergio Ranaldo}
\ead{sergio.ranaldo@poliba.it}

\author[dimeg]{Giuseppe Pascazio}
\ead{giuseppe.pascazio@poliba.it}

\author[dimeg]{Marco D. de Tullio}
\ead{marcodonato.detullio@poliba.it}

\cortext[cor]{Corresponding author}

\address[unibas]{Scuola di Ingegneria, Universit\`{a} degli Studi  della Basilicata, Viale dell'Ateneo Lucano 10 -- 85100 Potenza, Italy}

\address[dimeg]{Dipartimento di Meccanica, Matematica e Management, Politecnico di Bari, Via Re David 200 -- 70125 Bari, Italy}

\begin{abstract}
In this work, a dynamic-Immersed--Boundary method combined with a BGK-Lattice--Boltzmann technique is developed and critically discussed. The fluid evolution is obtained on a three-dimensional lattice with 19 reticular velocities (D3Q19 computational molecule) while the immersed body surface is modeled as a collection of Lagrangian points responding to an elastic potential and a bending resistance. A moving least squares reconstruction is used to accurately interpolate flow quantities and the forcing field needed to enforce the boundary condition on immersed bodies. The proposed model is widely validated against well known benchmark data for rigid and deformable objects. Rigid transport is validated by computing the settling of a sphere under gravity for five different conditions. Then, the tumbling of inertial particles with different shape is considered, recovering the Jefferey orbit for a prolate spheroid. Moreover, the revolution period for an oblate spheroid and for a disk-like particle is obtained as a function of the Reynolds number. The existence of a critical Reynolds number is demonstrated for both cases above which revolution is inhibited. The transport of deformable objects is also considered. The steady deformation of a membrane under shear for three different mechanical stiffness is assessed. Then, the tumbling  of a weakly-deformable spheroid under shear is systematically analyzed as a function strain stiffness, bending resistance and membrane mass.
\end{abstract}
\end{frontmatter}

\section*{Introduction}

By definition a capsule is composed of a membrane, with prescribed permeability and mechanical properties, enclosing some internal medium and controlling the exchange (e.g., fluid and momentum) between the environment and the enclosed medium~\cite{barthes2009,barthesARFM2016}. 
Examples of capsules can be found in nature (cells, bacteria, eggs, eye-bulbs) as well as in the man-made universe (medicines, cosmetics, food- and oil-industries). Specifically, the transport of micro- and nano-membranes within established flows represents one of the physical phenomena of great relevance for a wide range of engineering applications. The scientific community is investing its efforts in general and reliable numerical models able to describe micro- and nano-structures navigating complex geometries, such as: transport of cells and biological particles in the circulatory system and in \textit{in-vitro} devices~\cite{freundARFM2014,sebastian-dittrichARFM2018,secombARFM2017,guglietta2020}, systemic injection of drugs~\cite{fedosov2014,coclite20181,coclite20183,gentile2008,giverso2010,kruger2012}; rheology of active colloidal suspensions~\cite{bialke2015,nazockdast2016,wittkowski2017}; bacteria, micro-swimmers and micro-robots~\cite{laugaARFM2016,buzhardt2019,stricker2017,theers2018,nitti2020}; microfluidics devices for oil industrial applications with fixed or variable geometries~\cite{patel2017,lai2008,maiti2007,sun2016}. Indeed, all of these applications require the the detailed description of the complex mechanism driving such particles' transport. Reliable and accurate fluid-mechanical models are thus fundamental for the rational design and optimization of capsules abilities.

In this perspective, the Immersed Boundary (IB) method emerged a very attractive technique for the treatment of rigid and deformable structures immersed into the fluid stream~\cite{peskin2002}. Within this method, the fluid and the structure solutions are obtained and coupled together at each time step via a transfer function. Specifically, the fluid evolution is obtained on an Eulerian grid not conformed to the immersed object, the latter being represented by a collection of Lagrangian markers responding to some constitutive models. The choice of the adopted constitutive models is essentially arbitrary as well as the forcing procedure (either kinematic or dynamic) transferring boundary conditions imposed by the presence of the body onto the neighboring Eulerian fluid cells~\cite{coclite20191}. 
Recognizing its absolute flexibility, several scientists employed the IB technique adopting different constitutive models, fluid/structure coupling strategies, and forcing procedures for the fluid boundary conditions.~\cite{derosis2014,ye2014,kruger2011,melchionna2011,zhang2004,MDdTJCP2016}. 

In this work, a dynamic-Immersed--Boundary method is combined with a BGK-Lattice-- Boltzmann (BGK-LB) technique for describing the evolution of rigid and soft inertial capsules transportded in incompressible flows. This method was extensively studied and validated by the authors for two-dimensional problems~\cite{coclite20183,coclite20163,coclite20172,coclite20194}. Here, the authors propose an extension of such scheme to three-dimensional manifolds. Specifically, the BGK-LB equation is discretized and solved onto a three-dimensional lattice with 19 reticular velocities (D3Q19 computational molecule) describing the fluid evolution with second order accuracy. External boundaries are treated by imposing Dirichlet boundary conditions for the disctribution functions as formulated by Zou-He~\cite{zouhe1997}. On the other hand, the argument by Guo et al.~\cite{guo2011} for the projection of the forcing term onto the D3Q19 computational molecule is employed to enforce the immersed structures boundary conditions into the flow field. The Moving Least Squares (MLS) reconstruction~\cite{liu2005} is used to accurately interpolate the velocity and forcing fields between the Eulerian and Lagrangian points. Both rigid bodies and soft membranes are considered. The former ones are transported by solving the Euler-Newton equations for the center of mass and the revolution angles; the latter ones are transported as a collection of Lagrangian points, connected by means of a triangular surface mesh, responding to a bending resistance, an elastic potential, and the enclosed volume conservation constraint.   

The proposed model is validated against well known benchmark data for rigid and deformable capsules. First, the second order accuracy of the present formulation is demonstrated through a grid-refinement study with five different Eulerian meshes on a lid-driven square cavity. Then, the position and velocity of a sphere settling under gravity are computed and compared against the experimental findings by Ten Cate et al.~\cite{tencate2002}. The falling dynamics is analyzed for different values of Reynolds number, Froude number and solid/fluid density ratio. The tumbling motion of inertial rigid particles with different shape is also considered. On one side, the Jefferey orbit of a prolate spheroid is recovered for small values of the Reynolds number (Re); while, one the other side, the revolution period for an oblate spheroid and for a disk-like particle is obtained as a function of Re and the existence of a critical value is demonstrated above which rotation is inhibited. Furthermore, the validation of the deformable particle constitutive model is carried out by computing the steady deformation of an elastic capsule under shear~\cite{lac2004}. 
The tumbling period of a weakly-deformable prolate spheroid under shear is systematically analyzed as a function of: strain stiffness, regulated by the non-dimensional shear rate (G = 0.01, 0.025, 0.05, 0.075, 0.1); bending resistance (Eb = 0.001, 0.005, 0.01); and membrane mass ($\rho s/\rho f=$ 1, 1.2, 1.5, 2, 5). Interestingly, the revolution period seems to be only influenced by the bending resistance. On the contrary, membrane stiffness and mass are responsible for the larger or smaller surface relative strain without affecting the revolution period.

\section{Method}

\subsection{\textsc{The 3D-BGK lattice-Boltzmann method}}
 
The fluid evolution is modeled on a 3D computational molecule in the velocity space with 19 directions, the D3Q19 lattice. These reticular directions are identified by 19 vectors in $\mathbb{R}^3$ corresponding to discrete velocities ${\bf e_i}({\bf x})$ ($i=0,\ldots,18$), where ${\bf x}$ are the spatial coordinates, whose Cartesian coordinates are the columns of the following matrix:\\
\begin{small}
\begin{equation}
{\bf e} = \begin{pmatrix}
0& 1&-1& 0& 0& 0& 0& 1& 1& 1& 1&-1&-1&-1&-1& 0& 0& 0& 0 \\
0& 0& 0& 1&-1& 0& 0& 1&-1& 0& 0& 1&-1& 0& 0& 1& 1&-1&-1 \\
0& 0& 0& 0& 0& 1&-1& 0& 0& 1&-1& 0& 0& 1&-1& 1&-1& 1&-1 \\
\end{pmatrix}
\label{e_i}
\end{equation}
\end{small}
On this cubic lattice, 19 discrete distribution functions $\{f_i\}$ ($i=0,\ldots,18$) are transported responding to the forced Boltzmann equation,
\begin{equation}
\label{evoleq}
{f_{i}({\bf x}+{\bf e}_{i}\Delta t, t+\Delta t)-f_{i}({\bf x}, t)=
-\frac{\Delta t}{\tau}[f_{i}({\bf x}, t)-f_{i}^{eq}({\bf x}, t)] +\Delta t {\cal F}_i}\, ,
\end{equation}
with $\Delta t$ the time step, $\tau$ the unique relaxation
time in the BGK-approximation~\cite{bgk}, $\{f^{eq}_i\}$ ($i=0,\ldots,18$) the local equilibrium distribution functions, and ${\cal F}_i$ ($i=0,\ldots,18$) the components of the forcing term accounting for the body boundary condition. 
The kinematic viscosity of the flow is related to $\tau$ as $\nu=c_s^2(\tau -\frac{\Delta t}{2})$ being $c_s=1/\sqrt{3}$ the reticular sound speed. The moments of the distribution functions define the fluid density $\rho=\sum_i f_i$, the momentum $\rho {\bf u}=\sum_i f_i {\bf e}_i+\frac{\Delta t}{2}{\bf f}_{lb}$ and the pressure $p= c_s^2 \sum_i f_i = c_s^2 \rho$~\cite{kruger2016}.\\ 
The local equilibrium distribution functions $\{f_i^{eq}\}$ ($i=0,\ldots,18$) are chosen to be representative of the Maxwell-Boltzmann distribution projected onto the lattice as follows:
\begin{equation}
\label{eqfunc}
f_i^{eq}({\bf x},t)=\omega_i\rho\left[1+\frac{1}{c_s^2}({\bf e}_i \cdot{\bf u})+
\frac{1}{2 c_s^4}({\bf e}_i \cdot{\bf u})^2-\frac{1}{2c_s^2}{\bf u}^2
\right]\, ,
\end{equation}
with $\omega_0=1/3$, $\omega_{1-6}=1/18$, and $\omega_{7-18}=1/36$.\\
Following the argument by Guo et al.~\cite{guo2002}, ${\cal F}_i$ reads:
\begin{equation}
\label{forcing}
{{\cal F}_i= \Bigl(1-\frac{1}{2\, \tau}\Bigr)\omega_i\Bigl[\frac{{\bf e_i}-{\bf u}}{c_s^2}+\frac{{\bf e_i}\cdot{\bf u}}{c_s^4}{\bf e_i}\Bigr]\cdot {\bf f}_{lb}}\, 
\end{equation}
where ${\bf f}_{lb}({\bf x},t)$ is a the body force term accounting for the presence of immersed objects. Dirichlet boundary conditions are imposed on external boundaries treated with the known--velocity bounce back procedure by Zou and He~\cite{zouhe1997}. It is proved that in such a framework one can recover the forced Navier--Stokes equation with second order accuracy~\cite{junk2009,junk2011}.

Although, for each $\tau$ in the stability range $\tau > \frac{1}{2}$ the scheme converge to the same solution, higher value of $\tau$ lead to higher value of the reference velocity $u_{ref}$ and, consequently, higher non-dimensional time interval $L_{ref}/u_{ref}$ corresponding to a single time step ($L_{ref}$ being the reference length)~\cite{kruger2016}. As the matter of fact, $\tau$ regulates the viscosity and $\nu$ regulates $u_{ref}$ through the Reynolds number Re = $\frac{u_{ref} L_{ref}}{\nu}$. Due to these reasons, all of the simulations in the present work are computed with $\tau = 0.75$.

\subsection{\textsc{Immersed Boundary treatment}}

\begin{figure}[h]
\centering
\includegraphics[scale=0.27]{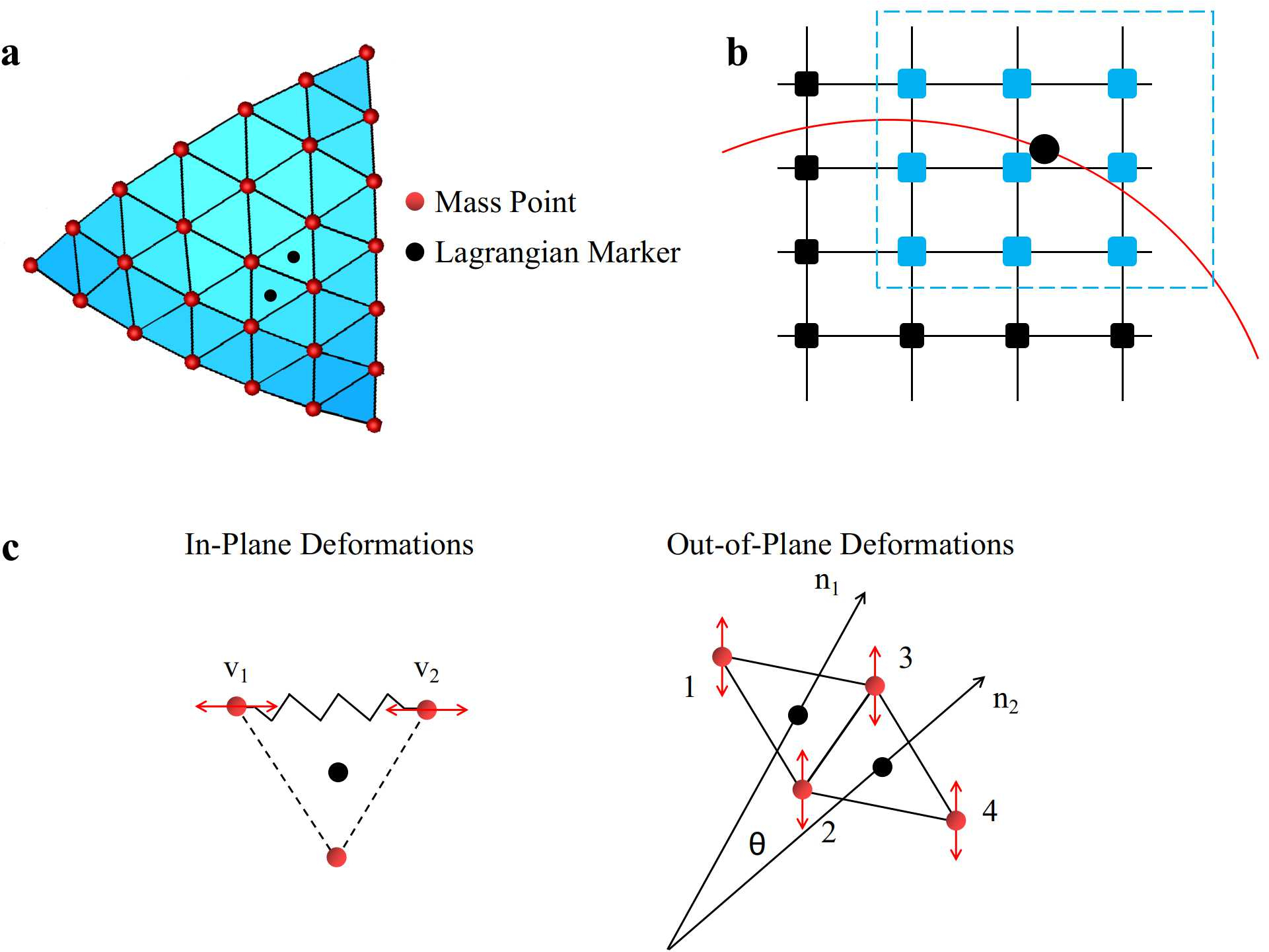}
\caption{{\bf Immersed-boundary technique schematic.} {\bf a.} Geometry description by means of triangular elements, whose centroids coincide with the Lagrangian markers. {\bf b.} Definition of the support domain (blue dashed square) centered on the selected Lagrangian marker (black circle) and the nine Eulerian points (per plane) involved in the forcing (blue squares). {\bf c.} Sketch of the two nodes (vertices of an edge) involved in the elastic spring model. Right: four nodes (belonging to the two triangles sharing an edge) involved in the bending spring model.}
\label{IB}
\end{figure}

In the Immersed-Boundary technique~\cite{peskin2002}, an obstacle in the flow can be considered as a collection of Lagrangian markers superimposed to the Eulerian withstanding fluid lattice ({\bf Figure.\ref{IB}a}). Here, the surface of the immersed geometry is discretized by means of a triangulated mesh, the Lagrangian markers coinciding with the centroids of each triangular element. The Lagrangian grid and Eulerian lattice interact through a transfer function built on a moving--least-squares (MLS) reconstruction~\cite{liu2005}, exchanging the distribution functions, while the body force term in Eq.\eqref{forcing}, ${\bf f}_{ib}$, is evaluated through the formulation by Favier et al.~\cite{pinelli2014}. 
Given a Lagrangian marker (with index $l$), 27 Eulerian points are identified, falling into the support domain, defined as a cube with side equal to $r_w=1.3\, \Delta x$ and centered on the Lagrangian marker (see {\bf Figure.\ref{IB}b}). Given the solution at time level $n$, the velocity component of Lagrangian marker \textit{l} is evaluated as,
\begin{equation}
\label{uLag}
{U_i({\bf x})=\sum_{k=1}^{27}\, \phi^l_k({\bf x}) u^k_i}\, ,
\end{equation}
where $u^k_i$ indicates the  $i$-th component of the velocity at the $k$-th Eulerian point associated with the marker $l$ and $\phi$ is the transfer operator obtained minimizing with respect to ${\bf a(x)}$ the following weighted L$_2$-norm:
\begin{equation}
\label{l2norm}
{J=\sum_{k=1}^{27} W({\bf x}-{\bf x}^k}) [{\bf p}^T({\bf x}^k){\bf a(x)}-u^k_i]^2\, .
\end{equation}
In the above equation, ${\bf p}^T$ is a basis function vector and ${\bf a(x)}$ a vector of coefficients such that 
$\sum_{k=1}^{27} \, \phi^l_k({\bf x}) u^k_i\, ={\bf p}^T({\bf x}^k){\bf a(x)}$, ${\bf x}$ the Lagrangian marker position, and $W({\bf x}-{\bf x}^k)$ a weight function.
In this work, a linear basis function, ${\bf p}^T=(1,x,y,z)$, is considered along with an exponential weight function:
\begin{equation}
\label{intKernel}
W({\bf x}-{\bf x}^k)= 
\begin{cases} 
e^{-(r_k/\alpha)^2}\, , &\mbox{if } r_k\leq 1 \\ 
0\, , & \mbox{if } r_k> 1\ \, , \\
\end{cases} 
\end{equation} 
where, $\alpha=0.3$ and $r_k$ is the distance between the Lagrangian point and the associated $k$-th Eulerian point normalized over the size of the support domain, $r_w$, $r_k=\frac{|{\bf x}-{\bf x}^k|}{r_w}$.
For each Lagrangian marker the volume force ${\bf F}_l$ required to impose the boundary condition can be evaluated and then 
transferred to the Eulerian points into the support domain using the same functions used before.
Being ${\bf U}_{b}^l({\bf x})$ the desired velocity at the boundary related to the $l$-th triangle, the Lagrangian volume force is:
\begin{equation}
{{\bf F}_l {\bf (x)}=\frac{{\bf U}_{b}^{l} {\bf (x)}-{\bf U(x)}}{\Delta t}}\, ,
\end{equation}
and then the body force for the $k$-th Eulerian point is finally,
\begin{equation}
\label{bodyforce}
{{\bf f}_{lb}}^k=\sum_{l} c_l \phi^l_k {\bf F}_l \, ,
\end{equation}
where the sum is extended to all Lagrangian markers having the $k$-th point in their support domain.
The scale coefficient $c_l$ is conveniently obtained by imposing the conservation between the Lagrangian and Eulerian grids of the total force acting on the fluid~\cite{vanella2009}.
The proposed algorithm concludes by inserting ${\bf f}_{ib}$ in equation~\eqref{forcing} and starting the next step by transporting the distribution functions ${\bf f}_i$ through equation~\eqref{evoleq}. In this way, the information given by the forcing term evaluated at the time level $n$ is transferred to the $n+1$-th time level. 

\subsection{\textsc{Fluid-Structure Interaction}}

Immersed bodies motion is determined by \textit{dynamics IB} technique described in \cite{coclite20191}, solving the Newton equation for each mass point, accounting for both internal and external stresses, namely, elastic boundary response and hydrodynamics forces, respectively. Here, a weak coupling approach is adopted.~\cite{coclite20163}

\textbf{\textit{Network model for deforming bodies.}} Immersed bodies internal stresses are computed as elastic strain response, bending resistance, and enclosed volume conservation (if applicable). The stretching elastic potential acting on the two vertices sharing the $l$-th edge ({\bf Figure.\ref{IB}c}) is given as
\begin{equation}
{V_{l}^{s}=\frac{1}{2}k_{s,l}(l_{l}-l_{l,0})^{2}}\, ,
\label{strainPot}
\end{equation}
being $k_{s,l}$ [N/m] the elastic constant associated to the edge $l$, $l_{l}$ its current length, and $l_{l,0}$ its length in the stress-free configuration. The nodal forces corresponding to the elastic energy for vertices 1 and 2 connected by \textit{l} reads:
\begin{equation}
{\begin{cases}
{\bf F}_{1}^{int,s}=-k_{s,l}(l-l_{0})\frac{{\bf r}_{1,2}}{l}\, , \\
{\bf F}_{2}^{int,s}=-k_{s,l}(l-l_{0})\frac{{\bf r}_{2,1}}{l}\, , \\
\end{cases}}
\label{strainFor}
\end{equation}
where ${\bf r}_{i,j}={\bf r}_{i}-{\bf r}_{j}$ and ${\bf r}_{i}$ is the position vector of \textit{i} with respect to \textit{j}. Here, the elastic constant for the $l$-th edge is computed trough the capillary number Ca$=\frac{\rho \nu u_{ref}}{k_{s,0}}$ and then distributed with a surface-weighted-averaging process (Van Gelder model~\cite{gelder}), $k_{s,l}=\frac{k_{s,0} T \sum_i A^l_i}{l^2}$. 
In the above relationships $\rho$ and $\nu$ correspond to surrounding fluid density and kinematic viscosity while $u_{ref}$ is a reference velocity, $k_{s,0}$ the body global elastic constant measured in [N/m], $T$ the immersed body thickness and $A^l_i$ the area of the two triangular elements sharing the $l$-th edge.

The out-of-plane deformation of two elements sharing an edge is computed as a bending resistance involving the four vertices of this couple of triangles. Given the initial (stress-free) and current local curvature, $k_{v,0}$ and $k_{v}$, respectively, the bending potential reads:
\begin{equation}
{V_{v}^{b}=\frac{1}{2}k_{b}(k_{v}-k_{v,0})^{2}}\, ,
\label{bendPot}
\end{equation}
with $k_{b}$ the bending constant. The local curvature is computed by measuring the variation of the angle between the unit normal vectors of two adjacent elements, ${\bf n}_1$ and ${\bf n}_2$, ($\theta -\theta _{0}$), with $\theta _{0}$ the angle in the stress free configuration ({\bf Figure.\ref{IB}c}). The bending constant associated to the two selected triangles is related to $k_{s,l}$ through the flexural rigidity $Eb=\frac{k_{b}}{k_{s,l}r^2}$.~~\cite{tan2012}  
The resulting forces acting on the four nodes read:
\begin{equation}
{\begin{cases}
{\bf F}^{int,b}_1 = \beta_b \bigl[
b_{11}({\bf n}_1 \times {\bf r}_{32}) + 
b_{12}({\bf n}_2 \times {\bf r}_{32}) \bigr] \, \\

{\bf F}^{int,b}_2 = \beta_b \bigl[
b_{11}({\bf n}_1 \times {\bf r}_{13}) + 
b_{12}({\bf n}_1 \times {\bf r}_{34}  + {\bf n}_2 \times {\bf r}_{13}) +
b_{22}({\bf n}_2 \times {\bf r}_{34}) \bigr] \, \\

{\bf F}^{int,b}_3 = \beta_b \bigl[
b_{11}({\bf n}_1 \times {\bf r}_{21}) + 
b_{12}({\bf n}_1 \times {\bf r}_{42}  + {\bf n}_2 \times {\bf r}_{21}) +
b_{22}({\bf n}_2 \times {\bf r}_{42}) \bigr] \, \\

{\bf F}^{int,b}_4 = \beta_b \bigl[
b_{11}({\bf n}_1 \times {\bf r}_{23}) + 
b_{22}({\bf n}_2 \times {\bf r}_{23}) \bigr] \, \\
\end{cases}}\, ,
\label{bendFor}
\end{equation}
with $b_{ij}= \frac{{\bf n}_i\cdot {\bf n}_j}{|{\bf n}_i||{\bf n}_j|}$ and $\beta_b=k_b\frac{sin(\theta)cos(\theta_0)-cos(\theta)sin(\theta_0)}{\sqrt{1-cos^2(\theta)}}$.

The fluid volume enclosed by of the incompressible membrane is constrained with a penalty force acting on the i$^{th}$-element centroid:
\begin{equation}
{{\bf F}_{i}^{int,v}=-k_{v}\left(1-\frac{V}{V_{0}}\right) {\bf n}_{i} A_{i}}\, . 
\label{volFor}
\end{equation}
Here, $k_v$ is the incompressibility constant, $V$ and $V_0$ the current and initial enclosed volume and $A_i$ the current area of the $i$-th element.

\textbf{\textit{Hydrodynamics Stresses.}} Pressure and viscous stresses exerted by the fluid on the $i$-th triangular element are:
\begin{equation}
{{\bf F}_{i}^{ext,p}(t)=-(p^+_{i}-p^-_{i}) \, {\bf n}^+_{i} \, A_{i}}\, , \\
\label{hydroForP}
\end{equation}
\begin{equation}
{{\bf F}_{i}^{ext,\tau}(t)=(\bar{\bf \tau}^+_{i}-\bar{\bf \tau}^-_{i})\cdot {\bf n}^+_{i} \, A_{i}}\, ,
\label{hydroForNu}
\end{equation}
where $\bar{\bf \tau}^+_{i}$, $\bar{\bf \tau}^-_{i}$ and $p^+_{i}$, $p^-_{i}$ are the viscous stress tensor and the pressure on the centroid of the $i$-th element taken from the external (+) and internal fluid (-), respectively. ${\bf n^+}_{i}$ is the normal outward unit verctor. The pressure and velocity derivatives in Eq.s \eqref{hydroForP} and \eqref{hydroForNu} are evaluated using a probe in the normal positive and negative direction of each element, being the probe length $1.2\, \Delta x$, and using the presented moving least squares recontruction~\cite{vanella2009}. In this framework, the velocity derivatives evaluated at the probe are considered equal to the ones on the triangular element centroid and all force contributions are computed with respect to the centroid of each elements and then transferred to the vertices.~\cite{MDdTJCP2016}

\textbf{\textit{Elastic Membranes.}} The total force ${\bf F}_{v}^{tot}(t)$ acting on the $v$-th element of the immersed body is evaluated in time and the position of the vertices is updated at each time step considering the membrane mass uniformly distributed over the vertices,
\begin{equation}
{m_{v}\dot{{\bf u}}_{v}={\bf F}_{v}^{tot}(t)={\bf F}_{v}^{int,s}(t)+{\bf F}_{v}^{int,b}(t)+{\bf F}_{v}^{int,v}(t)+{\bf F}_{v}^{ext,p}(t)+{\bf F}_{v}^{ext,\tau}(t)}\, ,
\label{newtonSoft}
\end{equation}
where $m_v$ and $\dot{{\bf u}}_{v}$ indicate the mass and acceleration of vertex $v$, respectively.

The Newton equation of motion is integrated with the Verlet algorithm using, as first tentative velocity, the value obtained by interpolating the fluid velocity from the surrounding lattice nodes, 
${\bf u}_{v,0}(t+\Delta t)$, 
\begin{equation}
{{\bf x}_{v}(t+\Delta t)=
{\bf x}_{v}(t)
+{\bf u}_{v,0}(t+\Delta t)
\Delta t+\frac{1}{2}\frac{{\bf F}_{v}^{tot}(t)}{m_{v}}\Delta t^{2}+O(\Delta t^{3})}\, ;
\label{verletPos}
\end{equation}
then, the velocity at the time level $t+\Delta t$ is computed as:
\begin{equation}
{{\bf u}_{v}(t+\Delta t)=\frac{\frac{3}{2}{\bf x}_{v}(t+\Delta t)-2{\bf x}_{v}(t)+\frac{1}{2}{\bf x}_{v}(t-\Delta t)}{\Delta t}+O(\Delta t^{2})}\, .
\label{verletVel}
\end{equation}
Finally, the position at the time level $t+\Delta t$ is computed employing again eq. (\ref{verletPos}) with the velocity obtained by eq. (\ref{verletVel}).

\textbf{\textit{Rigid objects.}} Rigid motion is readily obtained integrating all stress contributions, eqs. (\ref{hydroForP}) and (\ref{hydroForNu}),
over the boundary of the immersed body and then updating both linear and angular velocity: ${\bf F}^{tot}(t)=m\, \dot{{\bf u}}(t)$ and ${\bf M}^{tot}(t)= {\bf I}\, \dot{{\bf \omega}}(t)$. Here $m$ is the particle mass, ${\bf F}^{tot}(t)$ is the total force exerted by the fluid on the particle; ${\bf M}^{tot}(t)$ is the total moment acting on the rigid body, and ${\bf I}$ is the tensor of inertia. Finally, ${\bf u}(t)$ and ${\bf \omega} (t)$ are computed as:
\begin{equation}
{{\bf u}(t)=\frac{2}{3}(2{\bf u}(t-\Delta t)-\frac{1}{2}{\bf u}(t-2\Delta t)+\dot{{\bf u}}(t)\Delta t)+O(\Delta t^{2})}\, ,
\label{linearVelRig}
\end{equation}
\begin{equation}
{{\bf \omega} (t)=\frac{2}{3}(2{\bf \omega} (t-\Delta t)-\frac{1}{2}{\bf \omega} 
(t-2\Delta t)+\dot{{\bf \omega}}(t)\Delta t)+O(\Delta t^{2})}\, .
\label{angularVelRig}
\end{equation} 

\subsection{\textsc{Parallel Computing Framework}}

Recently, a number of parallelization techniques tailored on the BGK--LB scheme have been proposed~\cite{korner2006,lb3d2017,woodgate2018}. The D3Q19 computational molecule is composed of a set of nine points per plane arranged along three computational rows thus resulting in a square with side length equal to 2 $\Delta x$. Due to this compact nature of the D3Q19 computational molecule, one of the most suitable choice is to split the computational lattice into Cartesian blocks and entrust the solution of Eq.\eqref{evoleq} to different threads; each thread is appointed for the computational nodes within its boundaries and for its communicating boundary planes (back and forward) to adjacent threads. This logic is readily implemented by using the Message Passing Interface (MPI) Cartesian CPU-grid generator function with dimensions $N^{cpu}_{x},\, N^{cpu}_{y},\, N^{cpu}_{z}$ in the three spatial directions, respectively. So that, the total number of threads evaluating the computational lattice is given by $N^{cpu}_{x} \times N^{cpu}_{y} \times N^{cpu}_{z}$.
At each time step the distribution functions corresponding to boundaries (either planes, edges or angular points) between blocks are transferred from one to the adjacent block forward and backward in the three coordinate directions. Two ghost-node lines are defined for each spatial direction of each Cartesian block to have that the support domain of a Lagrangian marker falling into a block would be completely defined within the same block. The immersed structure dynamics is handled using the MPI Cartesian block subdivision, each CPU computing the dynamics of the Lagrangian markers falling into its boundaries. Specifically, each thread reads the immersed structure local positions and velocities while only the thread owing the selected Lagrangian marker computes its internal and external stresses thus resulting in non--null values of the forces defined in Eq.s~\eqref{strainFor}, \eqref{bendFor}, \eqref{volFor}, \eqref{hydroForP}, and \eqref{hydroForNu}. In this way the forces exerted by the structure derive from the sum of those calculated by all of the CPUs. Indeed, the sum is done through an MPI collective communicator.

\section{Results and discussion}

\subsection{\textsc{Scheme validation and accuracy: Lid-driven cavity test case}}

\begin{figure}
\centering
\includegraphics[scale=0.275]{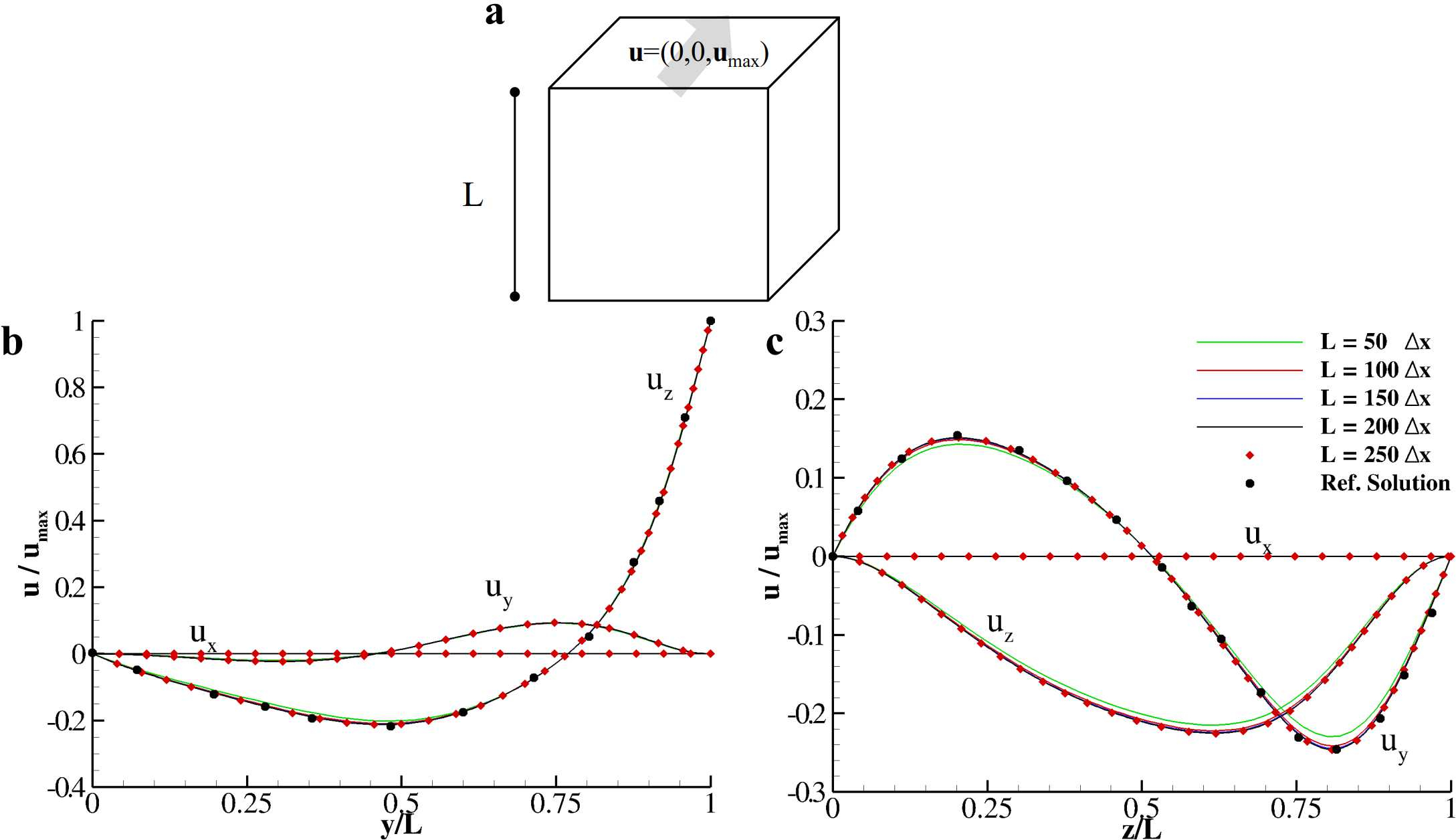}
\caption{{\bf Lid-driven cavity.} {\bf a.} Sketch of the physical problem. {\bf (b,c)} Comparison between the distributions of $u_x$, $u_y$, and $u_z$ taken at $x=z=0.5L$ ({\bf b}) and $x=y=0.5L$ ({\bf c}) and the benchmark data by Napolitano and Pascazio~\cite{pascazio1991} for the lid-driven cavity at Re=100 computed with L=50, 100, 150, 200 and 250 $\Delta x$.}
\label{LidCavity}
\end{figure}
\begin{figure}
\centering
\includegraphics[scale=0.25]{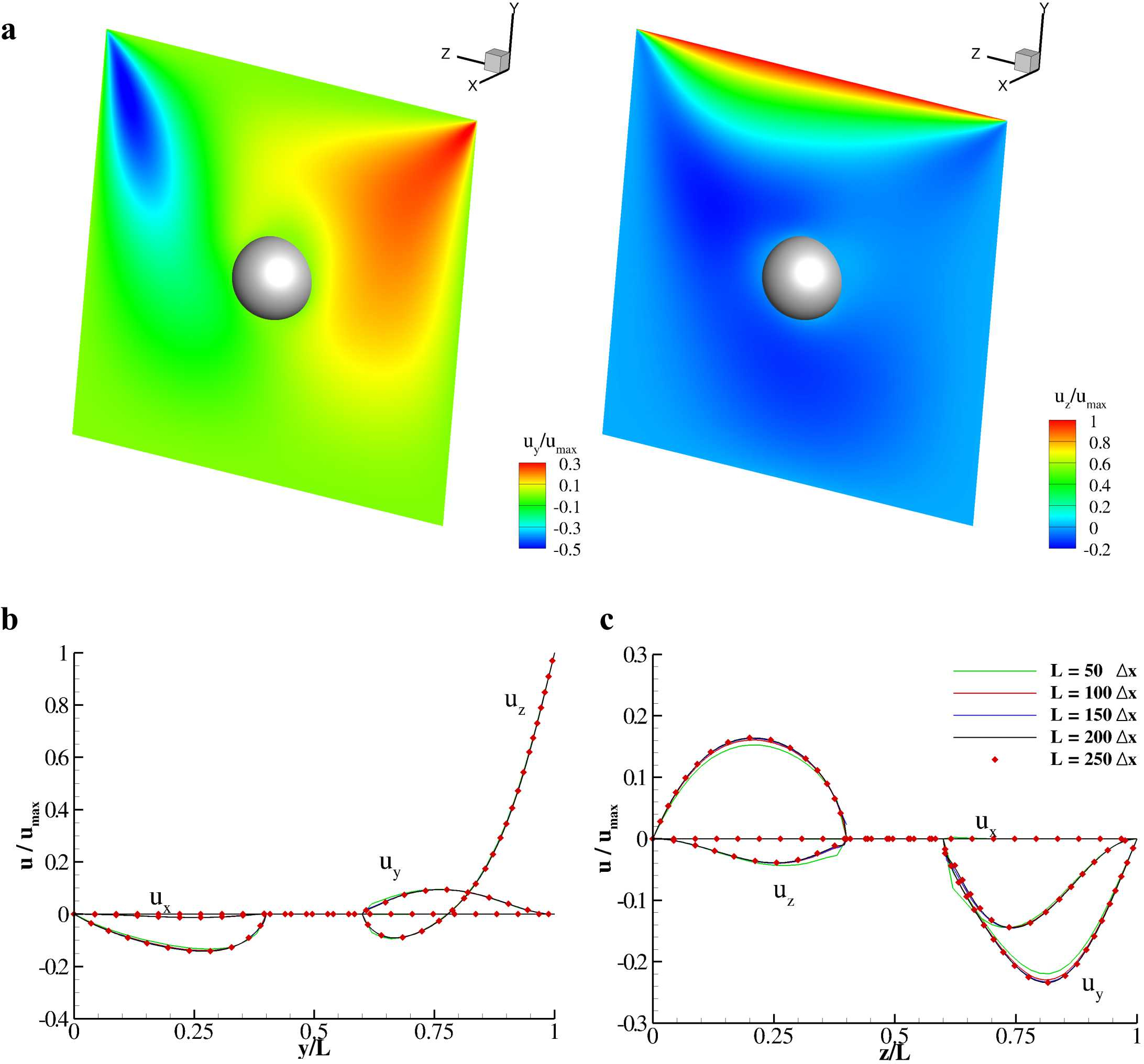}
\caption{{\bf Lid-driven cavity with a spherical obstacle immersed in.} {\bf a.} Contour plot of $u_y$ (left) and $u_z$ (right) taken at $x=0.5\, L$ for L = 250 $\Delta x$. {\bf (b,c)}  Distribution of $u_x$, $u_y$, and $u_z$ taken at $x=z=0.5L$ ({\bf b}) and $x=y=0.5L$ ({\bf c}) for the lid-driven cavity at Re=100 computed with L=50, 100, 150, 200 and 250 $\Delta x$ with a spherical obstacle in the center by diameter $d=0.2\, L$.}
\label{LidCavitySphere}
\end{figure}
\begin{figure}
\centering
\includegraphics[scale=0.275]{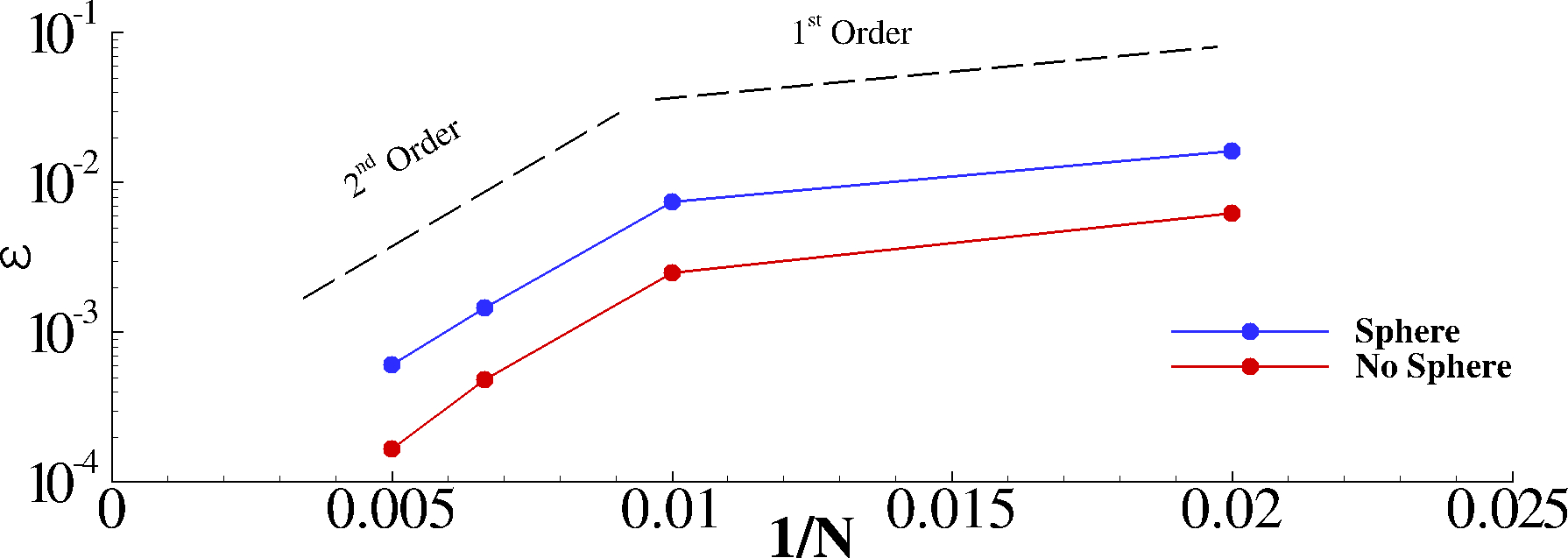}
\caption{{\bf Grid refinement study.} Mesh-refinement study on $u_z$ distributions for the lid-driven cavity with and without the spherical obstacle immersed in.}
\label{LidCavityErr}
\end{figure}

To validate the computational method for the fluid evolution and determine its accuracy a lid-driven cavity test case is computed at first. Specifically, a box with side L is considered, discretized with five meshes, containing 50, 100, 150, 200 and 250 $\Delta x$ cells. The wall at $y=y_{max}$ (see {\bf Figure.\ref{LidCavity}a}) moves with velocity ${\bf u}_w=(0,0,u_{max})$. The upper wall velocity is regulated by the Reynolds number Re (=$\frac{u_{max} L}{\nu}$), $\nu$ being the fluid kinematic viscosity, related to the relaxation parameter $\tau$ equal to one for all of these computations. The resulting velocity distributions taken at  x=z=0.5 L ({\bf Figure.\ref{LidCavity}b}) and x=y=0.5 L ({\bf Figure.\ref{LidCavity}c}) are compared with the benchmark solutions by Napolitano and Pascazio~\cite{pascazio1991} thus validating the quality of the present predictions. In particular, for the distribution at  x=z=0.5 L, the largest difference for $u_z$ with respect to the reference solution is found for L=50 $\Delta x$ at 0.48\, L, resulting in a relative error of 0.015. 
For the distribution taken at x=y=0.5 L, the largest difference for $u_y$ with respect to the reference solution is found for L=50 $\Delta x$ at 0.75\, L and the relative error is 0.018. The $L^2$-norm of the $u_z$ relative error with respect to the reference solution (obtained with L=250 $\Delta x$), $\epsilon=||\frac{u-u_{L=250\Delta x}}{u_{L=250\Delta x}}||^2$, is depicted in {\bf Figure.\ref{LidCavityErr}}. The distribution of $\epsilon$ as a function the mesh size (parameterized with 1/N) shows the second-order accuracy of the method, with the error already largely smaller than 1$\%$ for L=50 $\Delta x$. 
Moreover, in order to analyze the overall scheme accuracy within the coupling technique between fluid and the immersed boundary, a solid spherical obstacle with diameter 0.2\, L is placed in the center of the computational domain (see {\bf Figure.\ref{LidCavitySphere}a}). The solid body is discretized with a number of triangular elements corresponding to about five Lagrangian points for each Eulerian cell. The computed flow field is reported in {\bf Figure.\ref{LidCavitySphere}a} depicting the velocity components along z and y axis for the reference Eulerian grid L = 250 $\Delta x$. Moreover, the distributions of $u_x$, $u_y$, and $u_z$ taken at x=z=0.5 L and x=y=0.5 L are documented for all of the five computed Eulerian grids in {\bf Figure.\ref{LidCavitySphere}b} and {\bf Figure.\ref{LidCavitySphere}c}, respectively. The $L^2$-norm of the $u_z$ relative errors depicts that overall, while preserving the second-order space accuracy, scheme errors are slightly higher when considering the coupled fluid-structure system although well below O(10$^{-2}$) when N$\ge 100$ (see {\bf Figure.\ref{LidCavityErr}}).

\subsection{\textsc{Rigid spheres settling under gravity}}

\begin{figure}
\centering
\includegraphics[scale=0.275]{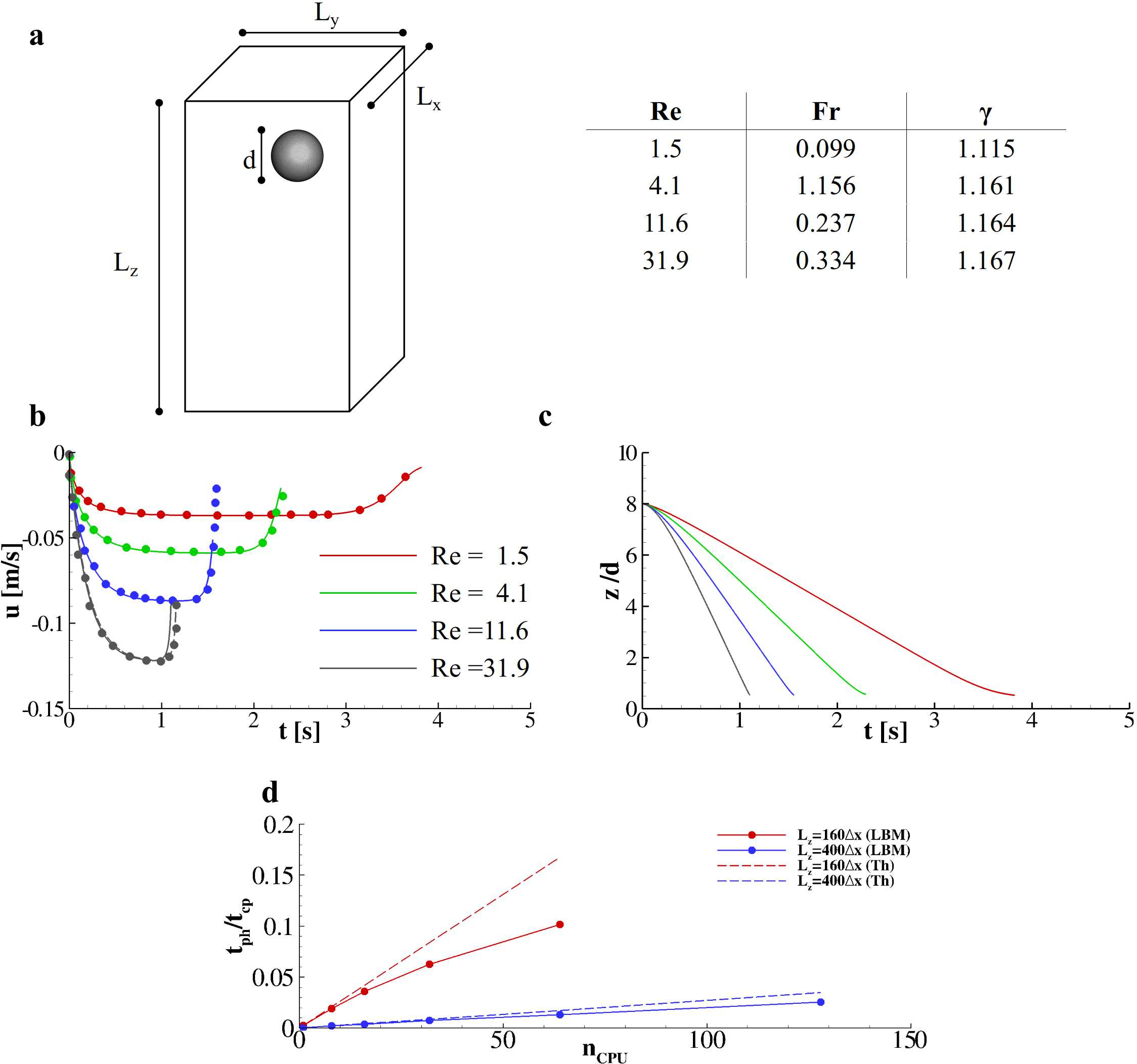}
\caption{{\bf Rigid spheres settling under gravity.} {\bf a.} Sketch of the physical problem with characteristic dimensions and non-dimensional groups used for the computations. {\bf b.} Comparison between the computed settling velocity distributions and benchmark experimental data by Ten Cate et al~\cite{tencate2002}. The dashed line corresponds to the velocity distribution obtained with d = 50 $\Delta x$. {\bf c.} Distribution of the particles' centroid positions obtained during the settling process. {\bf d.} Measure of the speedup performances computed over two different computational meshes, namely with $L_z$ = 160 and 400 $\Delta x$ over different number of threads running in parallel. Solid lines refer to present model computations while dashed lines correspond to theoretical scaling.}
\label{TenCate}
\end{figure}

The simulation of a rigid sphere settling under gravity in a square cylinder filled with fluid at rest is presented here against the experiments performed by A. ten Cate et al.~\cite{tencate2002} used as benchmark. These experiments analyzed both centroid vertical velocity and position using Particle Image Velocimetry (PIV) technique, thus providing accurate measures of the falling spheres from the initial condition at rest until the stop at the bottom of the channel.
The computational domain is a box with size $(L_x,L_y,L_z)=(6.67\, d, 6.67\, d, 10.67\, d)$, $d$ corresponding to the diameter of the settling sphere discretized here with 30 $\Delta x$. The particle is a rigid plain sphere placed in the center of the channel at $L_{p,z}=8\, d$; its surface is described by a triangular mesh of about 10'000 elements. As previously analyzed by the authors, the ratio between the Lagrangian mesh and the Eulerian grid characteristic lengths cannot be arbitrary into this specific immersed boundary framework~\cite{coclite20191,coclite20163}. A convenient choice would be to keep this ratio between 0.3 and 0.5 corresponding to three-to-five Lagrangian points per each Eulerian cell. The sphere density is fixed by the ratio between the solid and the fluid density $\gamma = \rho_s/\rho_f$ while the Reynolds number (Re=$\frac{d u_{inf}}{\nu}$) regulate the fluid viscosity $\nu$ and is based on the settling velocity $u_{inf}$ obtained by the relation for the drag coefficient $C_d$ by Abraham~\cite{abraham1970}:
\begin{equation}
{C_d= C_0\left(1+\frac{\delta_0}{\sqrt{Re}}\right)^2}\, ,
\end{equation}
where $C_0\delta_0^2=24$ and $\delta_0=9.06$ and reads,
\begin{equation}
{u_{inf}=\sqrt{\frac{4gd}{3C_d}(\gamma-1)}}\, .
\end{equation}
Finally, the Froude number (Fr =$ \frac{u_{inf}}{\sqrt{gd}}$) returns the gravity acceleration $g$. The values for Re, Fr, and $\gamma$ used in the computations are reported in {\bf Figure.\ref{TenCate}a}. 
The agreement between present model predictions and the experimental data is depicted in term of the distributions of the particles' centroid velocities during the settling process ({\bf Figure.\ref{TenCate}b}). The relative error between the two set of data, based on the displacement achieved in the peak z-velocity, $\epsilon=\frac{max(|u_{LBM}|)-max(|u_{exp}|)}{max(|u_{exp}|)}$ ranges from 0.0023 to 0.0039 for Re = 1.5 and 31.9, respectively. For Re = 31.9 and Fr = 0.334 the ascendant part of the velocity distribution shows a slight displacement from the experimental data while the descendant part and the peak value are well reproduced. In fact, due to the increasing of both the Froude and the Reynolds numbers, the particle passes through multiple computational cells in a single Newtonian time step. The proposed model for enforcing the boundary conditions on the immersed body needs to gradually reconstruct the pressure field based on the corrections made on the velocity field avoiding any refilling procedure for solid-to-fluid cells transition~\cite{coclite20163}. This gives a slight inaccuracy into the ascendant part of the velocity distribution that is essentially determined by the restraining of the particle because of the pressure field near the southern wall~\cite{tencate2002}. By increasing the resolution of the Eulerian grid (d = 50 $\Delta x$) the pressure field is accurately reconstructed and the displacement between the two velocity distributions disappears (see the dashed line in {\bf Figure.\ref{TenCate}b}).

The trajectory of the settling spheres are shown in {\bf Figure.\ref{TenCate}c}. To assess the performance of the Cartesian parallel framework used and presented in the {\bf Method} section, the test case with the higher Re is re-computed on two different meshes with $L_z$ = 160 and 400 $\Delta x$ over a growing number of threads running in parallel. As depicted in {\bf Figure.\ref{TenCate}d}, for the larger mesh the weak-speedup performance is quite similar to the theoretical expectations while for the smaller mesh the communications between the CPUs dominate the wall time increasing the overall computational burden. Then, a strong-speedup test is performed by computing three different meshes and varying with the same growth factor the number of CPUs running in parallel for each Cartesian directions thus keeping constant the number of computational cells entrusted to each tread. This test returns similar wall times for the three meshes documented in {\bf Table.\ref{parallel}}.
\begin{table}
\centering
\begin{tabular}{ccc}
\toprule
CPU  & Points in $L_z$ & Total Wall Time \\
\midrule
$1^3$  & $100$  & $1680\, s $ \\
$2^3$  & $200$  & $1721\, s $ \\        
$4^3$  & $400$  & $1778\, s $ \\
\bottomrule
\end{tabular}
\caption{Parallel performance for different lattice dimensions and number of CPUs used.}
\label{parallel}
\end{table}

\subsection{\textsc{Tumbling motion of rigid spheroids under shear}}
      
\begin{figure}
\centering
\includegraphics[scale=0.275]{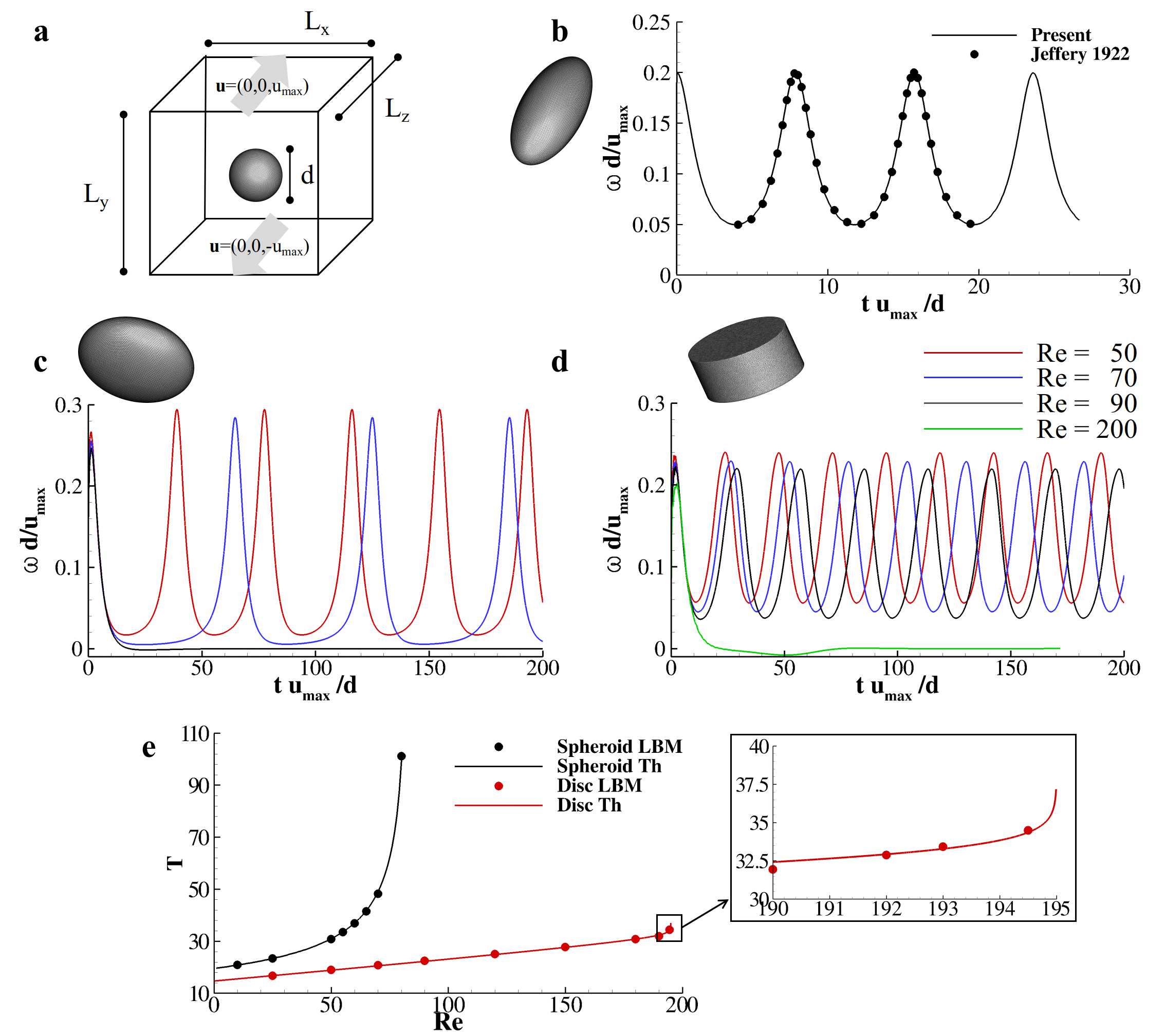}
\caption{{\bf Rigid spheroids under shear.} {\bf a.} Sketch of the physical problem. {\bf b.} Comparison between the angular velocity obtained with the analytical solution by Jefferey~\cite{jeffery1922} and present model predictions for a prolate plain spheroid freely moving under shear. {\bf (c,d)} Angular velocity distribution for an oblate spheroid ({\bf c}) and a discoidal particle ({\bf d}) under shear obtained at different Reynolds number. {\bf e.} Revolution period as a function of the Reynolds number for an oblate spheroid and a discoidal particle. The solid lines correspond to the analytical predictions.}
\label{Spheroid}
\end{figure}      

To further validate the proposed model, the revolution of a plain neutrally--buoiant rigid spheroid in a shear flow established by two countermoving walls is studied ({\bf Figure.\ref{Spheroid}a}). 
First, the revolution period of an initially resting prolate spheroid placed in the center of a cubic domain is computed for a shear flow at Re = 10$^{-2}$. The particle Reynolds number follows the definition by Huang et al~\cite{huang2012} and is based on the moving walls velocity, $u_{max}$, the particle major axis length, d, and the kinematic viscosity of the surrounding fluid, Re=$\frac{4\gamma d^2}{\nu}$ being $\gamma=\frac{u_{max}}{1/2L_y}$ the shear rate. 
Such particle is a spheroid with aspect ratio 2 (ratio of the major diameter over the two minor ones), the length of its major axis is d ($=60\Delta x$) and its surface is discretized with 20'500 triangular elements. The particle is placed with the major axis parallel to x in the center of the cubic box with side $L_x=L_y=L_z=4\, d$.
The distribution of the rotational velocity around x is compared with the analytical solution obtained by Jefferey~\cite{jeffery1922} based on the hypothesis of minimum dissipation energy thus finding an excellent agreement considering both the revolution period and the angular velocity distribution ({\bf Figure.\ref{Spheroid}b}). Specifically, the present model prediction for the revolution period is 7.853 against the analytical value of 7.854 while the largest displacement for $\omega d/u_{max}$ is found at 15.67 $t u_{max}/d$ returning a relative error of 0.000832.

Then, the revolution period of an oblate spheroid rotating under shear at increasing values of the Reynolds number is measured. This spheroid is an oblate particle with aspect ratio 2 (ratio of the two major diameters over the  minor one), d corresponding to the length of the two major diameters discretized with 60 $\Delta x$, again the particle surface is discretized with 20'500 triangles. Initially at the rest, the particle is placed in the center of a box with dimension $L_x=2.5\, d$, $L_y=5.0\, d$, and $L_z=6.8\, $. Ding and Aidun~\cite{aidun1998} and Zettner and Yoda~\cite{zettner2001} found the existence of a critical Reynolds number (Re$_c$) for the angular tumbling velocity; for $Re>Re_c$ the particle stops its revolution motion assuming a fixed position with the plane containing the major axis forming an angle $\theta_f$ with respect to z direction.
Precisely, the tumbling period T$_t$ as a function of the Reynolds number is predicted by the scaling law, $T_t=c(Re_c-Re)^{-1/2}$. Huang et al~\cite{huang2012} observed that for an oblate particle with aspect ratio 2 the critical Reynolds number is 80 and c = 180 while Ding and Aidun~\cite{aidun1998} found Re$_c=81$ and $c=200$. In the present formulation the tumbling angular velocity recovers the described phenomenology, as demonstrated in {\bf Figure.\ref{Spheroid}c}; the scaling law is recovered by using $c=179.5$ and Re$_c=81.3$ (see {\bf Figure.\ref{Spheroid}e}) and $\theta_f= 0.29\, \pi$ for $Re=90$.
Lastly, the tumbling motion of a circular cylinder with aspect ratio 2 (ratio between the circle diameter d and its height) is analyzed. Specifically, the discoidal particle presents a diameter d corresponding to 60 $\Delta x$ and its surface covered by about 30'000 triangles. Interestingly, the discoidal particle tumbling presents a critical Reynolds number of 192.5; above this value the revolution of the disc does not occur, as depicted in {\bf Figure.\ref{Spheroid}d}, and $\theta_f= 0.16\, \pi$ for $Re=200$. The distribution of tumbling period as a function of the Reynolds number is documented in {\bf Figure.\ref{Spheroid}e}. Such distribution does not follow a power law in the form $T_t=c(Re_c-Re)^{-1/2}$ as seen in the oblate spheroid tumbling. However, a scaling law may be established in the form $T_t=c[(A-Re)(Re_c-Re)^{-\frac{1}{50}}]$ with $c = \frac{1}{10}$, A = 182, and Re$_c$ = 195.

\subsection{\textsc{Deforming capsules under linear laminar flow}}

\begin{figure}
\centering
\includegraphics[scale=0.275]{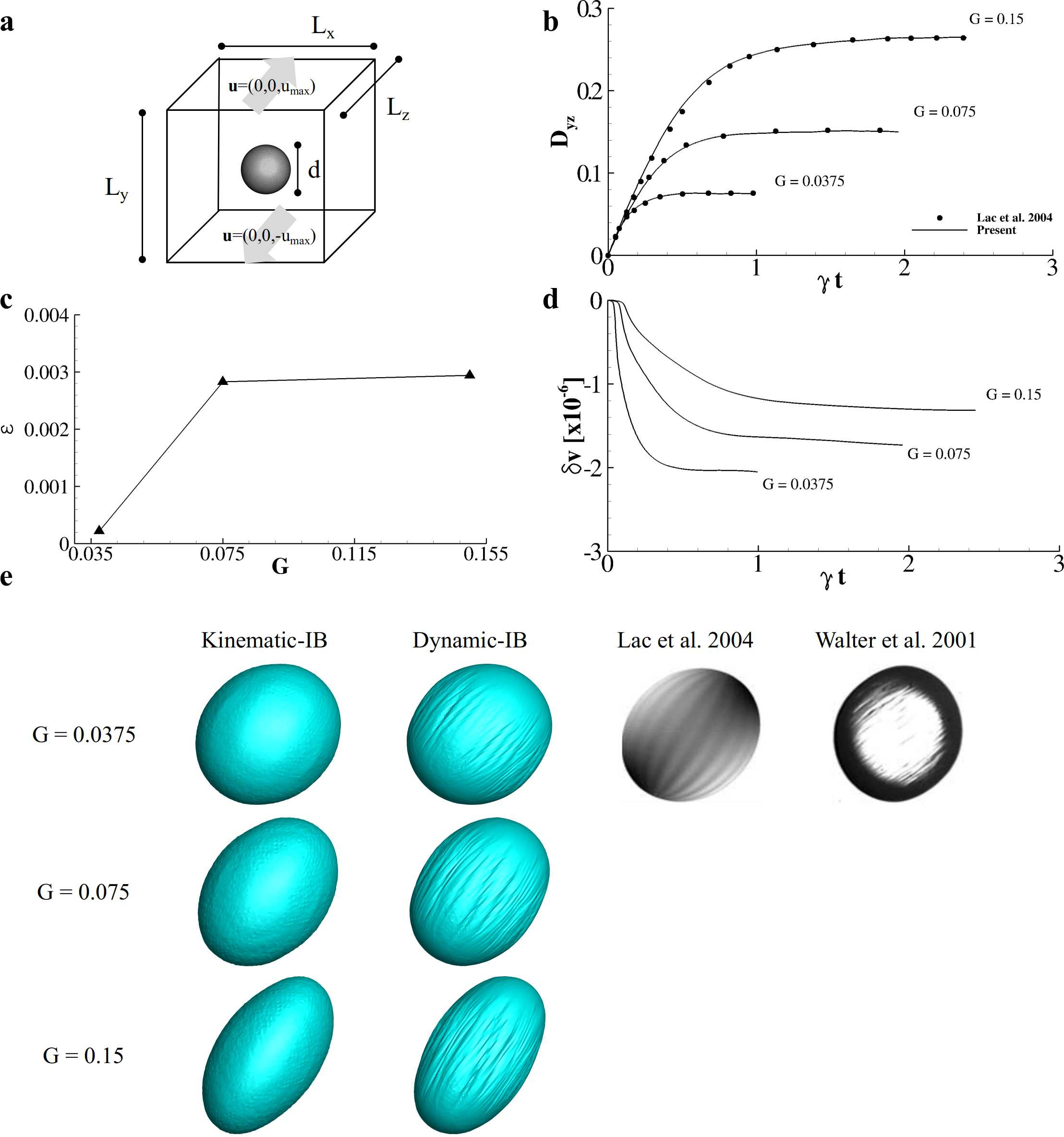}
\caption{{\bf Capsules deforming under shear.} {\bf a.} Schematic representation of a spherical capsule deforming under linear laminar flow. {\bf b} Comparison between the distributions of the Taylor parameter over time obtained for G = 0.0375, 0.075, and 0.15 with the proposed model and the benchmark data by Lac et al.~\cite{lac2004}. {\bf c} Relative error obtained for the three investigated values of G with respect to the benchmark data by Lac et al.~\cite{lac2004}. {\bf d} Variation of the capsule relative volume over time as a function of G. {\bf e} Configurations of the steady-state deformed capsules for G = 0.0375, 0.075, and 0.15 predicted by the proposed Dynamic-IB scheme as compared to the Kinematic-IB scheme and to the computations by Lac et al.~\cite{lac2004} as well as with the experimental data by Walter et al~\cite{walter2001}.}
\label{Capsule}
\end{figure}   

The dynamics of the shear-induced deformation of an initially spherical capsule is studied in this section. Referring to {\bf Figure.\ref{Capsule}a}, a spherical capsule with diameter d is placed at the center of a cubic box with side length H = 10 d. The capsule diameter is discretized by 10 $\Delta x$ while the capsule surface is discretized with $n_L =$ 10000 triangular elements. The imposed shear rate is $\gamma = \frac{u_{max}}{H/2}$, with $u_{max}$ the velocity magnitude of the top and bottom walls; the resulting Reynolds number equals 0.01 ($=\frac{\gamma (d/2)^2}{\nu}$). The mechanical behavior of the capsule is regulated by the elastic constant $k_s$, computed using the dimensionless shear rate G $= \frac{\nu \rho \gamma d/2}{k_s}$. The mass of the capsule is defined considering a unitary solid-to-fluid density ratio, $\frac{\rho s}{\rho f}=1$, being $\rho s$ and $\rho f$ the solid and fluid densities, respectively; the volume constraint constant is chosen as $k_v = 10$.

The quality of the present model predictions is measured against well known published data used as benchmark~\cite{lac2004}. Specifically, the evolution of the Taylor deformation parameter in the y-z plane, D$_{yz}$, at x = 0.5 H is analyzed, D$_{yz}$ being defined as the weighted ratio between the major (M) and the minor (m) diameter in the selected plane, namely D$_{yz}=\frac{M-m}{M+m}$. The distribution of D$_{yz}$ for G = 0.0375, 0.075, and 0.15 is documented in {\bf Figure.\ref{Capsule}b} against the data by Lac et al.~\cite{lac2004}. An excellent agreement is found for all of the three values of the shear rate. The relative error with respect to benchmark data ($\epsilon = \frac{D_{yz} - D_{yz,Lac}}{D_{yz,Lac}}$) is smaller than 1\% in all of the computations (see {\bf Figure.\ref{Capsule}c}). Moreover, the distribution of the relative variation of the enclosed volume ($\delta v = \frac{v-v_0}{v_0}$) with respect to the undeformed configuration is depicted in {\bf Figure.\ref{Capsule}d}. Overall, $\delta v$ is well confined within the order of magnitude of 10$^{-6}$. 

As previously observed by Walter et al~\cite{walter2001} for an organosiloxane membrane, a capsule with a nearly null Poisson ratio immersed in a simple shear flow will present corrugations near its equator region due to the compression generated as response to shear-induced stretching. Specifically, the membrane section laying on the vorticity plane will stretch finding the steady equilibrium between hydrodynamics and internal stresses; while the off-vorticity section responds to this stretching with a compression. This mechanism brings to a phenomenon largely studied in literature known as membrane buckling~\cite{li2008,knoche2011,barthes2009}. 
This shape-instability has a numerical origin, consistent with the null bending resistance employed model, while the wrinkles location are physical insofar as they occur in the compressed portions of such membranes.~\cite{barthesARFM2016,dupont2015} Interestingly, the present dynamic-IB model  is able to reproduce the membrane buckling in agreement with the computations performed by Lac et al.~\cite{lac2004} and the experimental findings by Walter et al~\cite{walter2001} (see {\bf Figure.\ref{Capsule}e}). On the contrary, the kinematic-IB scheme, in which the immersed body is deformed advecting the Lagrangian points velocity with the fluid velocity, reads a smooth velocity distribution in all planes and directions and responds with a perfectly smooth membrane configuration ({\bf Figure.\ref{Capsule}e}). Note that the kinematic-IB and the dynamic-IB only differ by the technique used to displace Lagrangian points, as discussed by Coclite et al.~\cite{coclite20191}.    

\subsection{\textsc{Membrane tumbling motion as a function of the bending resistance}}

\begin{figure}
\centering
\includegraphics[scale=0.25]{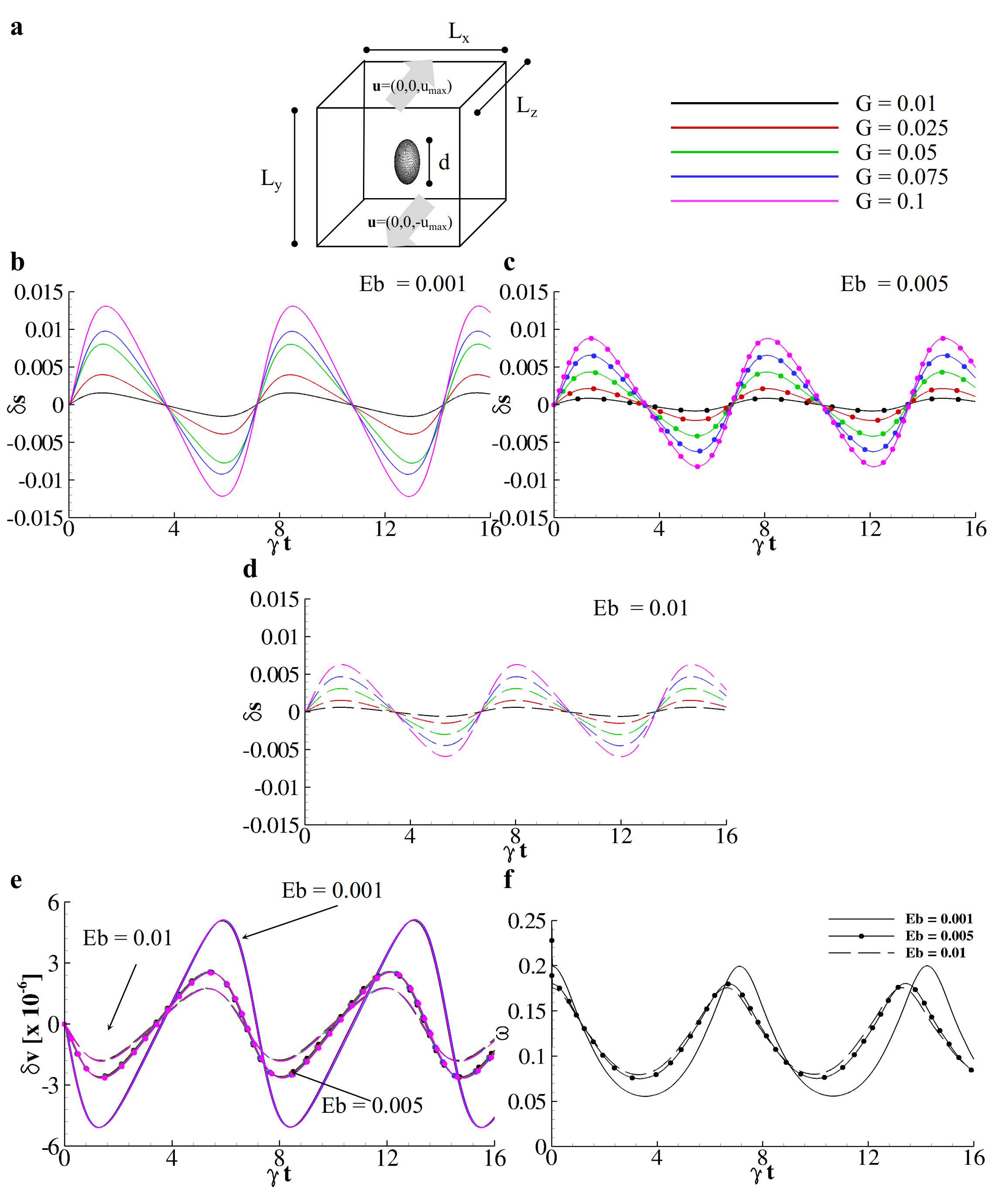}
\caption{{\bf Tumbling motion of a deformable prolate membrane.} {\bf a.} Schematic of the physical problem. ({\bf b},{\bf c},{\bf d}) Distribution of the relative variation of the membrane surface as a function of G for Eb = 0.001 ({\bf b}), 0.005 ({\bf c}), and 0.01 ({\bf d}). {\bf e.} Relative variation of the membrane volume for different values of G and Eb. {\bf f.} Membrane angular velocity over time for Eb = 0.001, 0.005, 0.01.}
\label{Tumbling_Gam1}
\end{figure}

The tumbling motion of an inertial membrane in a linear laminar flow is considered. As depicted in {\bf Figure.\ref{Tumbling_Gam1}a} a prolate capsule with major axis d is placed in the center of a cubic box with side length H = 10 d. The minor axis dimension of the capsule is equal to half the major axis length. d corresponds to 10 $\Delta x$ while the capsule surface is discretized with $n_L =$ 10000 triangular elements. The fluid boundary conditions are kept identical to the previous test case, so that: the imposed shear rate is $\gamma = \frac{u_{max}}{H/2}$, with $u_{max}$ the velocity magnitude of moving walls; the Reynolds number Re$=\frac{\gamma (d/2)^2}{\nu}$ = 0.01. The membrane elastic constant $k_s$ is computed by the dimensionless shear rate G $= \frac{\nu \rho \gamma d/2}{k_s}$. The bending resistance is related to the strain stiffness and is regulated by Eb$=\frac{k_{b}}{k_{s,l}(d/2)^2}$. The volume constraint constant is chosen as $k_v = 10$. The tumbling motion is systematically assessed in term of the membrane stiffness (G = 0.01, 0.025, 0.05, 0.075, and 0.1); the bending resistance, Eb = 0.001, 0.005, 0.01; and the solid/fluid density ratio, $\rho s/\rho f=$ 1, 1.2, 1.5, 2, and 5. 

\begin{figure}
\centering
\includegraphics[scale=0.275]{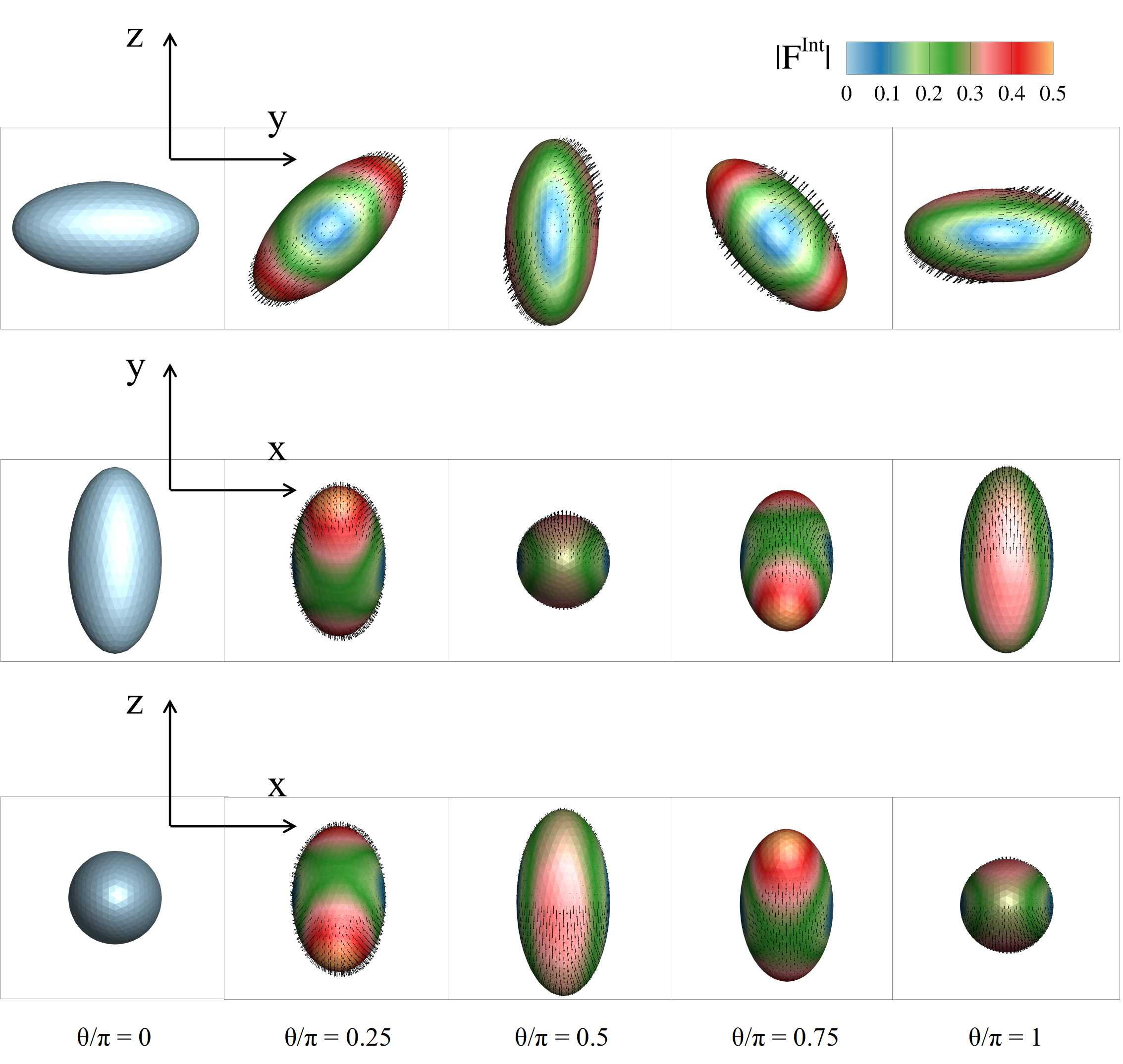}
\caption{{\bf Particle configurations.} Membrane configurations for G = 0.01 and Eb = 0.001 taken at $\theta / \pi =$ 0, 0.25, 0.5, 0.75, and 1 with superimposed the contour of the module of the internal forces vectors indicating the exerted hydrodynamics stresses.}
\label{Tumbling_Conf}
\end{figure}

First, a freely moving neutrally buoyant ($\rho s/\rho f=$ 1) membrane immersed in a linear shear flow is considered. As depicted in {\bf Figures.\ref{Tumbling_Gam1}b-d} such elastic capsule stretches in a periodic fashion with a period that depends on Eb. As clearly shown, the amplitude of the relative surface stretching $\delta s = \frac{s-s_0}{s_0}$ depends on the stiffness. Precisely, the peak surface strain grows with G while is retrained by higher values of Eb. Such membranes strain the most for $\gamma t =$ 1.40, 1.35, and 1.30 with an half period of 3.705, 3.472, and 3.419 for Eb = 0.001, 0.005, and 0.01, respectively; indeed, they are about in the unperturbed configuration (null relative surface deviation)  at $\gamma t =$ 3.705, 3.472, and 3.419. Interestingly, the distribution of the relative volume deviation $\delta v$ shows a different behavior. All of the distributions characterized by the same value of the bending resistance overlap, as demonstrated in {\bf Figures.\ref{Tumbling_Gam1}e}. Moreover, we computed the tumbling velocity in the yz plane around their body center. The distributions of $\omega=\omega_{LB} d/ u_{max}$ in time are reported in {\bf Figure.\ref{Tumbling_Gam1}f} and demonstrate that there is a specific revolution period for each value of the bending stiffness while within these values the variation of G is uninfluential. Morevoer, such membranes experience the largest strain for multiple of $\frac{\pi}{4}$ corresponding to the points with the larger angular acceleration. Five configurations of the capsule revolution are sketched in {\bf Figures.\ref{Tumbling_Conf}} for G = 0.01 and Eb = 0.001. Note that, the higher internal constitutive forces are registered far from the particle equator due to larger hydrodynamics stresses (vectors in {\bf Figures.\ref{Tumbling_Conf}}). On the contrary, in the equator region the membrane is responding to the hydrodynamics stresses with almost null internal forces.   

\begin{figure}
\centering
\includegraphics[scale=0.275]{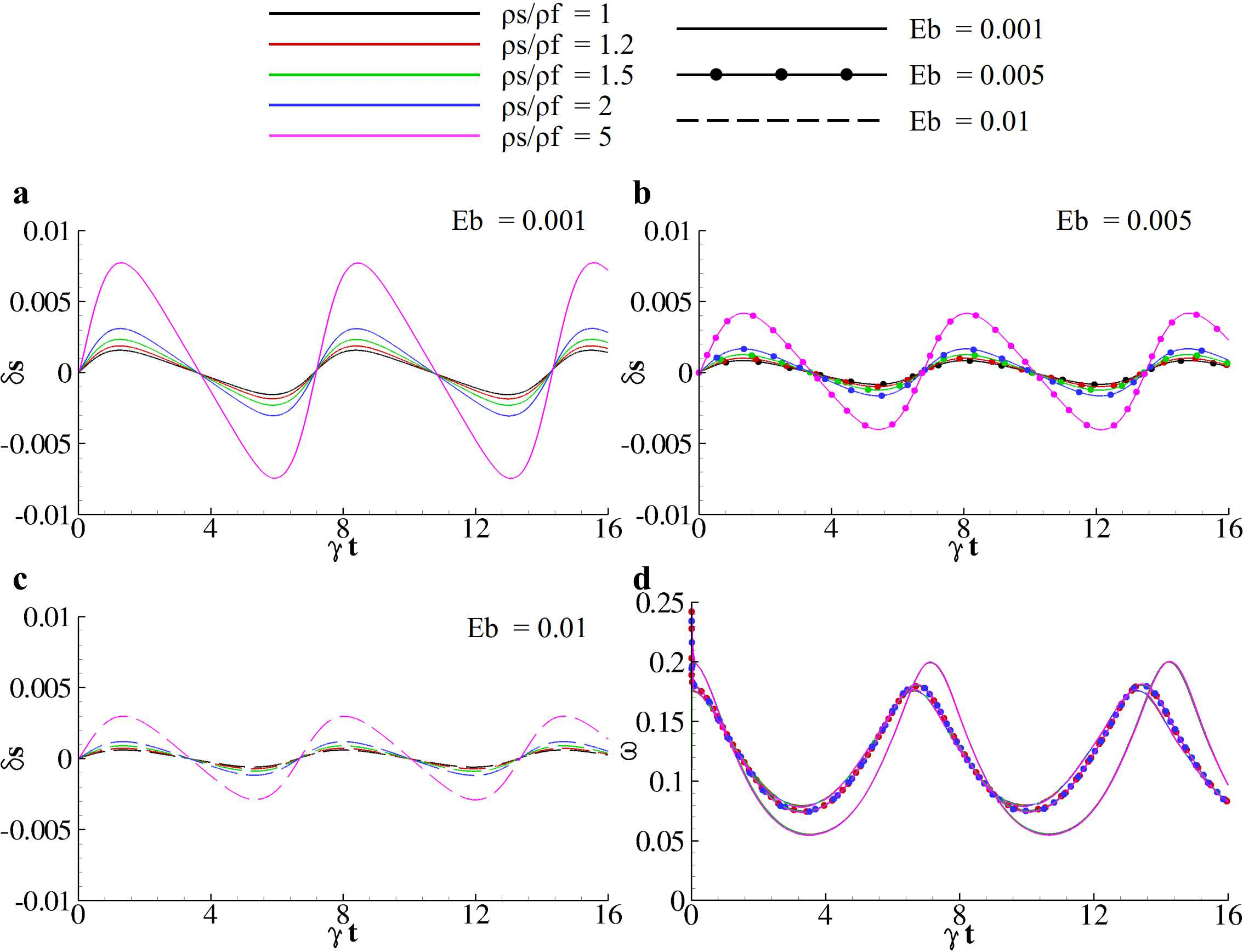}
\caption{{\bf Effect of the membrane mass on the tumbling period.} ({\bf a}, {\bf b}, {\bf c}) Distribution of the relative variation of the membrane surface for G = 0.01 and Eb = 0.001 ({\bf a}), 0.005 ({\bf b}), and 0.01 ({\bf c}) as a function of the membrane mass. {\bf b.} Membrane angular velocity for G = 0.01 and Eb = 0.001, 0.005, 0.01 obtained varying the membrane mass.}
\label{Tumbling_mass}
\end{figure}

Lastly, the effect of the membrane mass on the particle tumbling is considered by varying the ratio between the solid and the fluid densities for G = 0.01 and Eb = 0.001, 0.005, and 0.01. Specifically, the membrane mass is distributed considering a unitary thickness uniformly on the membrane surface. Five different values of the density ratio between the solid shell and the withstanding fluid are considered, namely $\rho_s/\rho_f =$ 1, 1.2, 1.5, 2, and 5. Indeed, higher values of the solid shell mass imply higher membranes inertia and consequently larger relative surface stretches (see {\bf Figure.\ref{Tumbling_mass}a-c}). Interestingly, for the investigated range of parameters the tumbling period is not affected by particles inertia and it is only dependent on the bending resistance coefficient. As the matter of fact, {\bf Figure.\ref{Tumbling_mass}d} demonstrates that particles with different inertia present the same angular velocity distribution in time.

\section*{Conclusions and future work}

A combined dynamic-Immersed Boundary (IB)--Lattice Boltzmann (LB) method is designed and employed originally to predict the interaction between low Reynolds number flows and deformable or rigid objects with arbitrary shape and buoyancy. A Moving Least Squares reconstruction was implemented to accurately interpolate the velocity and force fields between Eulerian grid describing the fluid evolution and the Lagrangian mesh defining the immersed object dynamics.

The quality of the present model predictions is measured by several validation tests for both rigid and deforming particles. Firstly, the accuracy of the present LB-IB scheme is computed by a grid-refinement study for the flow in a lid-driven square cavity with and without a spherical obstacle immersed in. This analysis demonstrates that the second order accuracy is conserved by the IB treatment. Then, the settling under gravity of a sphere is assessed for different values of the Reynolds and Froude numbers and sphere mass. Moreover, the tumbling motion of inertial rigid particles with different shape is considered. On one side, the Jefferey orbit of a prolate spheroid is recovered for a prolate ellipsoid; on the other hand, the existence of a critical value of the Reynolds number regulating the revolution period for an oblate spheroid and a disk rotating under shear is demonstrated. The dynamics of a spherical elastic capsule deforming under shear is computed and the obtained distribution of the Taylor parameter for different mechanical stiffness are validated against well known computations used as benchmark. Interestingly, the present model is able to capture the characteristic membrane buckling emerging in consolidated experiments and simulations for thin elastic capsules. Weakly deformable prolate spheroids tumbling under shear are then considered and their revolution period measured as a function of the mechanical stiffness, the bending resistance and the membrane mass. Interestingly, this systematic study revealed that, for the analyzed range of parameters, the revolution period is influenced by only the bending resistance. The membrane stretching stiffness and its mass are responsible for the larger or smaller surface relative strain but do not seem to play any role into the spheroids tumbling period itself.

Collectively, these data demonstrate the present formulation as a versatile and reliable framework for the microfluidics fluid/structure interaction problems. This approach can be readily applied to model the dynamics of biological micro-vesicles and particles for biomedical applications as well as to predict the transport of colloids into industrial microchips. The dynamic behavior of capsule containing a different fluid with respect to the suspension fluid is a challenging problem and will be the core of future investigations.

\section*{Acknowledgments}
This investigation has been partially supported by the Italian Ministry of Education, University and Research under the Programme “Department of Excellence” Legge 232/2016 (Grant No. CUP - D94I18000260001)

\section*{Competing Interests}
The authors declare no competing interests. 


\begin{thebibliography}{62}
\providecommand{\natexlab}[1]{#1}
\providecommand{\url}[1]{\texttt{#1}}
\expandafter\ifx\csname urlstyle\endcsname\relax
  \providecommand{\doi}[1]{doi: #1}\else
  \providecommand{\doi}{doi: \begingroup \urlstyle{rm}\Url}\fi

\bibitem[Barth{\`e}s-Biesel(2009)]{barthes2009}
Dominique Barth{\`e}s-Biesel.
\newblock Capsule motion in flow: Deformation and membrane buckling.
\newblock \emph{Comptes Rendus Physique}, 10\penalty0 (8):\penalty0 764--774,
  2009.

\bibitem[Barth{\`e}s-Biesel(2016)]{barthesARFM2016}
Dominique Barth{\`e}s-Biesel.
\newblock Motion and deformation of elastic capsules and vesicles in flow.
\newblock \emph{Ann. Rev. Fliud Mech.}, 48:\penalty0 25--52, 2016.

\bibitem[Freund(2014)]{freundARFM2014}
Jonathan~B. Freund.
\newblock Numerical simulation of flowing blood cells.
\newblock \emph{Ann. Rev. Fliud Mech.}, 46:\penalty0 67--95, 2014.

\bibitem[Sebastian and Dittrich(2018)]{sebastian-dittrichARFM2018}
Bernhard Sebastian and Petra~S. Dittrich.
\newblock Microfluidics to mimic blood flow in health and disease.
\newblock \emph{Ann. Rev. Fliud Mech.}, 50:\penalty0 483--504, 2018.

\bibitem[Secomb(2017)]{secombARFM2017}
Timothy~W. Secomb.
\newblock Blood flow in the microcirculation.
\newblock \emph{Ann. Rev. Fliud Mech.}, 49:\penalty0 443--461, 2017.

\bibitem[Guglietta et~al.(2020)Guglietta, Behr, Biferale, Falcucci, and
  Sbragaglia]{guglietta2020}
Fabio Guglietta, Marek Behr, Luca Biferale, Giacomo Falcucci, and Mauro
  Sbragaglia.
\newblock On the effects of membrane viscosity on transient red blood cell
  dynamics.
\newblock \emph{Soft Matter}, 2020.

\bibitem[M{\"u}ller et~al.(2014)M{\"u}ller, Fedosov, and Gompper]{fedosov2014}
Kathrin M{\"u}ller, Dmitry~A. Fedosov, and Gerhard Gompper.
\newblock Margination of micro- and nano-particles in blood flow and its effect
  on drug delivery.
\newblock \emph{Scientific Reports}, 4:\penalty0 4871 EP --, 05 2014.
\newblock URL \url{https://doi.org/10.1038/srep04871}.

\bibitem[Mollica et~al.(2018)Mollica, Coclite, Miali, Pereira, Paleari,
  Manneschi, DeCensi, and Decuzzi]{coclite20181}
Hilaria Mollica, Alessandro Coclite, Marco~E. Miali, Rui~C. Pereira, Laura
  Paleari, Chiara Manneschi, Andrea DeCensi, and Paolo Decuzzi.
\newblock Deciphering the relative contribution of vascular inflammation and
  blood rheology in metastatic spreading.
\newblock \emph{Biomicrofluidics}, 12\penalty0 (4):\penalty0 042205, 2018.
\newblock \doi{10.1063/1.5022879}.
\newblock URL \url{https://doi.org/10.1063/1.5022879}.

\bibitem[Coclite et~al.(2018)Coclite, Pascazio, de~Tullio, and
  Decuzzi]{coclite20183}
A.~Coclite, G.~Pascazio, M.D. de~Tullio, and P.~Decuzzi.
\newblock Predicting the vascular adhesion of deformable drug carriers in
  narrow capillaries traversed by blood cells.
\newblock \emph{Journal of Fluids and Structures}, 82:\penalty0 638 -- 650,
  2018.
\newblock ISSN 0889-9746.
\newblock \doi{https://doi.org/10.1016/j.jfluidstructs.2018.08.001}.
\newblock URL
  \url{http://www.sciencedirect.com/science/article/pii/S0889974618301944}.

\bibitem[F. et~al.(2008)F., M., and P.]{gentile2008}
Gentile F., Ferrari M., and Decuzzi P.
\newblock The transport of nanoparticles in blood vessels: The effect of vessel
  permeability and blood rheology.
\newblock \emph{Annals of Biomedical Engineering}, 36\penalty0 (2):\penalty0
  254--261, 2008.
\newblock \doi{10.1007/s10439-007-9423-6}.

\bibitem[Giverso et~al.(2010)Giverso, Scianna, Preziosi, Lo~Buono, and
  Funaro]{giverso2010}
C.~Giverso, M.~Scianna, L.~Preziosi, N.~Lo~Buono, and A.~Funaro.
\newblock Individual cell-based model for in-vitro mesothelial invasion of
  ovarian cancer.
\newblock \emph{Mathematical Modelling of Natural Phenomena}, 5\penalty0
  (1):\penalty0 203--223, 2010.
\newblock \doi{10.1051/mmnp/20105109}.

\bibitem[Kr{\"u}ger(2012)]{kruger2012}
Timm Kr{\"u}ger.
\newblock \emph{Computer simulation study of collective phenomena in dense
  suspensions of red blood cells under shear}.
\newblock Springer Science \& Business Media, 2012.

\bibitem[Bialk{\'e} et~al.(2015)Bialk{\'e}, Speck, and L{\"o}wen]{bialke2015}
Julian Bialk{\'e}, Thomas Speck, and Hartmut L{\"o}wen.
\newblock Active colloidal suspensions: Clustering and phase behavior.
\newblock \emph{Journal of Non-Crystalline Solids}, 407:\penalty0 367--375,
  2015.

\bibitem[Nazockdast and Morris(2016)]{nazockdast2016}
Ehssan Nazockdast and Jeffrey~F Morris.
\newblock Active microrheology of colloidal suspensions: Simulation and
  microstructural theory.
\newblock \emph{Journal of Rheology}, 60\penalty0 (4):\penalty0 733--753, 2016.

\bibitem[Wittkowski et~al.(2017)Wittkowski, Stenhammar, and
  Cates]{wittkowski2017}
Raphael Wittkowski, Joakim Stenhammar, and Michael~E Cates.
\newblock Nonequilibrium dynamics of mixtures of active and passive colloidal
  particles.
\newblock \emph{New Journal of Physics}, 19\penalty0 (10):\penalty0 105003,
  2017.

\bibitem[Lauga(2016)]{laugaARFM2016}
Eric Lauga.
\newblock Bacterial hydrodynamics.
\newblock \emph{Ann. Rev. Fliud Mech.}, 48:\penalty0 105--130, 2016.

\bibitem[Buzhardt and Tallapragada(2019)]{buzhardt2019}
Jake Buzhardt and Phanindra Tallapragada.
\newblock Dynamics of groups of magnetically driven artificial microswimmers.
\newblock \emph{Physical Review E}, 100\penalty0 (3):\penalty0 033106, 2019.

\bibitem[Stricker(2017)]{stricker2017}
Laura Stricker.
\newblock Numerical simulation of artificial microswimmers driven by
  {M}arangoni flow.
\newblock \emph{Journal of Computational Physics}, 347:\penalty0 467--489,
  2017.

\bibitem[Theers et~al.(2018)Theers, Westphal, Qi, Winkler, and
  Gompper]{theers2018}
Mario Theers, Elmar Westphal, Kai Qi, Roland~G Winkler, and Gerhard Gompper.
\newblock Clustering of microswimmers: interplay of shape and hydrodynamics.
\newblock \emph{Soft matter}, 14\penalty0 (42):\penalty0 8590--8603, 2018.

\bibitem[Nitti et~al.(2020)Nitti, Kiendl, Reali, and de~Tullio]{nitti2020}
Alessandro Nitti, Josef Kiendl, Alessandro Reali, and Marco~D de~Tullio.
\newblock An immersed-boundary/isogeometric method for fluid--structure
  interaction involving thin shells.
\newblock \emph{Computer Methods in Applied Mechanics and Engineering},
  364:\penalty0 112977, 2020.

\bibitem[Patel et~al.(2017)Patel, Das, Kuipers, Padding, and Peters]{patel2017}
HV~Patel, S~Das, JAM Kuipers, JT~Padding, and EAJF Peters.
\newblock A coupled volume of fluid and immersed boundary method for simulating
  3d multiphase flows with contact line dynamics in complex geometries.
\newblock \emph{Chemical Engineering Science}, 166:\penalty0 28--41, 2017.

\bibitem[Lai et~al.(2008)Lai, Tseng, and Huang]{lai2008}
Ming-Chih Lai, Yu-Hau Tseng, and Huaxiong Huang.
\newblock An immersed boundary method for interfacial flows with insoluble
  surfactant.
\newblock \emph{Journal of Computational Physics}, 227\penalty0 (15):\penalty0
  7279--7293, 2008.

\bibitem[Maiti and Nigam(2007)]{maiti2007}
RN~Maiti and KDP Nigam.
\newblock Gas- liquid distributors for trickle-bed reactors: a review.
\newblock \emph{Industrial \& Engineering Chemistry Research}, 46\penalty0
  (19):\penalty0 6164--6182, 2007.

\bibitem[Sun and Sakai(2016)]{sun2016}
Xiaosong Sun and Mikio Sakai.
\newblock Numerical simulation of two-phase flows in complex geometries by
  using the volume-of-fluid/immersed-boundary method.
\newblock \emph{Chemical Engineering Science}, 139:\penalty0 221--240, 2016.

\bibitem[Peskin(2002)]{peskin2002}
Charles~S. Peskin.
\newblock The immersed boundary method.
\newblock \emph{Acta Numerica}, 11:\penalty0 479--517, 1 2002.
\newblock \doi{10.1017/S0962492902000077}.

\bibitem[Coclite et~al.(2019)Coclite, Ranaldo, de~Tullio, Decuzzi, and
  Pascazio]{coclite20191}
A.~Coclite, S.~Ranaldo, M.D. de~Tullio, P.~Decuzzi, and G.~Pascazio.
\newblock Kinematic and dynamic forcing strategies for predicting the transport
  of inertial capsules via a combined lattice {B}oltzmann immersed boundary
  method.
\newblock \emph{Computers {\&} Fluids}, 180:\penalty0 41--53, 2019.
\newblock ISSN 0045-7930.
\newblock \doi{https://doi.org/10.1016/j.compfluid.2018.12.014}.
\newblock URL
  \url{http://www.sciencedirect.com/science/article/pii/S0045793018304304}.

\bibitem[De Rosis et~al.(2014)De Rosis, Ubertini, and Ubertini]{derosis2014}
Alessandro De Rosis, Stefano Ubertini, and Francesco Ubertini.
\newblock A comparison between the interpolated bounce-back scheme and the
  immersed boundary method to treat solid boundary conditions for laminar flows
  in the lattice {B}oltzmann framework.
\newblock \emph{Journal of Scientific Computing}, 61\penalty0 (3):\penalty0
  477--489, 2014.
\newblock ISSN 1573-7691.
\newblock \doi{10.1007/s10915-014-9834-0}.
\newblock URL \url{http://dx.doi.org/10.1007/s10915-014-9834-0}.

\bibitem[Ye et~al.(2014)Ye, Ng, Tan, Leo, and Kim]{ye2014}
Swe~Soe Ye, Yan~Cheng Ng, Justin Tan, Hwa~Liang Leo, and Sangho Kim.
\newblock Two-dimensional strain-hardening membrane model for large deformation
  behavior of multiple red blood cells in high shear conditions.
\newblock \emph{Theoretical Biology and Medical Modelling}, 11\penalty0
  (1):\penalty0 19, 2014.

\bibitem[Krueger et~al.(2011)Krueger, Varnik, and Raabe]{kruger2011}
T.~Krueger, F.~Varnik, and D.~Raabe.
\newblock Efficient and accurate simulations of deformable particles immersed
  in a fluid using a combined immersed boundary lattice {B}oltzmann finite
  element method.
\newblock \emph{Computers and Mathematics with Applications}, 61\penalty0
  (12):\penalty0 3485 -- 3505, 2011.
\newblock ISSN 0898-1221.
\newblock \doi{http://dx.doi.org/10.1016/j.camwa.2010.03.057}.

\bibitem[Melchionna(2011)]{melchionna2011}
Simone Melchionna.
\newblock A model for red blood cells in simulations of large-scale blood
  flows.
\newblock \emph{Macromolecular Theory and Simulations}, 20\penalty0
  (7):\penalty0 548--561, 2011.

\bibitem[Zhang and Hisada(2004)]{zhang2004}
Q.~Zhang and T.~Hisada.
\newblock Studies of the strong coupling and weak coupling methods in {FSI}
  analysis.
\newblock \emph{International Journal for Numerical Methods in Engineering},
  60\penalty0 (12):\penalty0 2013--2029, 2004.
\newblock \doi{10.1002/nme.1034}.

\bibitem[de~Tullio and Pascazio(2016)]{MDdTJCP2016}
M.~D. de~Tullio and G.~Pascazio.
\newblock A moving-least-squares immersed boundary method for simulating the
  fluid-structure interaction of elastic bodies with arbitrary thickness.
\newblock \emph{Journal of Computational Physics}, submitted, 2016.

\bibitem[Coclite et~al.(2016)Coclite, de~Tullio, Pascazio, and
  Decuzzi]{coclite20163}
A.~Coclite, M.~D. de~Tullio, G.~Pascazio, and P.~Decuzzi.
\newblock A combined lattice {B}oltzmann and immersed boundary approach for
  predicting the vascular transport of differently shaped particles.
\newblock \emph{Computers {\&} Fluids}, 136:\penalty0 260 -- 271, 2016.
\newblock ISSN 0045-7930.
\newblock \doi{http://dx.doi.org/10.1016/j.compfluid.2016.06.014}.

\bibitem[Coclite et~al.(2017)Coclite, Mollica, Ranaldo, Pascazio, de~Tullio,
  and Decuzzi]{coclite20172}
A.~Coclite, H.~Mollica, S.~Ranaldo, G.~Pascazio, M.~D. de~Tullio, and
  P.~Decuzzi.
\newblock Predicting different adhesive regimens of circulating particles at
  blood capillary walls.
\newblock \emph{Microfluidics and Nanofluidics}, 21\penalty0 (11):\penalty0
  168, 2017.
\newblock ISSN 1613-4990.
\newblock \doi{10.1007/s10404-017-2003-7}.
\newblock URL \url{https://doi.org/10.1007/s10404-017-2003-7}.

\bibitem[Coclite and Gambaruto(2019)]{coclite20194}
Alessandro Coclite and Alberto~M. Gambaruto.
\newblock Injection of deformable capsules in a reservoir: A systematic
  analysis.
\newblock \emph{Fluids}, 4\penalty0 (3), 2019.
\newblock ISSN 2311-5521.
\newblock \doi{10.3390/fluids4030122}.
\newblock URL \url{https://www.mdpi.com/2311-5521/4/3/122}.

\bibitem[Zou and He(1997)]{zouhe1997}
Qisu Zou and Xiaoyi He.
\newblock On pressure and velocity boundary conditions for the lattice
  {B}oltzmann {BGK} model.
\newblock \emph{Physics of Fluids}, 9\penalty0 (6):\penalty0 1591--1598, 1997.
\newblock \doi{http://dx.doi.org/10.1063/1.869307}.

\bibitem[Guo et~al.(2011)Guo, Zheng, and Shi]{guo2011}
Zhaoli Guo, Chuguang Zheng, and Baochang Shi.
\newblock Force imbalance in lattice {B}oltzmann equation for two-phase flows.
\newblock \emph{Phys. Rev. E}, 83:\penalty0 036707, Mar 2011.
\newblock \doi{10.1103/PhysRevE.83.036707}.

\bibitem[Liu and Gu(2005)]{liu2005}
Gui-Rong Liu and Yuan-Tong Gu.
\newblock \emph{An introduction to meshfree methods and their programming}.
\newblock Springer Science \& Business Media, 2005.

\bibitem[Ten~Cate et~al.(2002)Ten~Cate, Nieuwstad, Derksen, and Van~den
  Akker]{tencate2002}
A~Ten~Cate, CH~Nieuwstad, JJ~Derksen, and HEA Van~den Akker.
\newblock Particle imaging velocimetry experiments and lattice-{B}oltzmann
  simulations on a single sphere settling under gravity.
\newblock \emph{Physics of Fluids}, 14\penalty0 (11):\penalty0 4012--4025,
  2002.

\bibitem[Lac et~al.(2004)Lac, Barthes-Biesel, Pelekasis, and
  Tsamopoulos]{lac2004}
E~Lac, D~Barthes-Biesel, NA~Pelekasis, and J~Tsamopoulos.
\newblock Spherical capsules in three-dimensional unbounded {S}tokes flows:
  effect of the membrane constitutive law and onset of buckling.
\newblock \emph{Journal of Fluid Mechanics}, 516:\penalty0 303--334, 2004.

\bibitem[Bhatnagar et~al.(1954)Bhatnagar, Gross, and Krook]{bgk}
P.~L. Bhatnagar, E.~P. Gross, and M.~Krook.
\newblock A model for collision processes in gases. i. small amplitude
  processes in charged and neutral one-component systems.
\newblock \emph{Phys. Rev.}, 94:\penalty0 511--525, May 1954.
\newblock \doi{10.1103/PhysRev.94.511}.

\bibitem[Krüger et~al.(2016)Krüger, Kusumaatmaja, Kuzmin, Shardt, Silva, and
  Viggen]{kruger2016}
Timm Krüger, Halim Kusumaatmaja, Alexander Kuzmin, Orest Shardt, Goncalo
  Silva, and Erlend~Magnus Viggen.
\newblock \emph{The Lattice Boltzmann Method - Principles and Practice}.
\newblock Springer, 10 2016.
\newblock ISBN 978-3-319-44647-9.
\newblock \doi{10.1007/978-3-319-44649-3}.

\bibitem[Guo et~al.(2002)Guo, Zheng, and Shi]{guo2002}
Zhaoli Guo, Chuguang Zheng, and Baochang Shi.
\newblock Discrete lattice effects on the forcing term in the lattice
  {B}oltzmann method.
\newblock \emph{Phys. Rev. E}, 65:\penalty0 046308, Apr 2002.
\newblock \doi{10.1103/PhysRevE.65.046308}.

\bibitem[Junk and Yang(2009)]{junk2009}
Michael Junk and Zhaoxia Yang.
\newblock Pressure boundary condition for the lattice boltzmann method.
\newblock \emph{Computers \& Mathematics with Applications}, 58\penalty0
  (5):\penalty0 922--929, 2009.

\bibitem[Junk and Yang(2011)]{junk2011}
Michael Junk and Zhaoxia Yang.
\newblock Asymptotic analysis of lattice boltzmann outflow treatments.
\newblock \emph{Communications in Computational Physics}, 9\penalty0
  (5):\penalty0 1117--1127, 2011.

\bibitem[Favier et~al.(2014)Favier, Revell, and Pinelli]{pinelli2014}
J.~Favier, A.~Revell, and A.~Pinelli.
\newblock A lattice {B}oltzmann-immersed boundary method to simulate the fluid
  interaction with moving and slender flexible objects.
\newblock \emph{Journal of Computational Physics}, 261:\penalty0 145--161,
  2014.
\newblock \doi{10.1016/j.jcp.2013.12.052}.

\bibitem[Vanella and Balaras(2009)]{vanella2009}
Marcos Vanella and Elias Balaras.
\newblock A moving-least-squares reconstruction for embedded-boundary
  formulations.
\newblock \emph{Journal of Computational Physics}, 228\penalty0 (18):\penalty0
  6617 -- 6628, 2009.
\newblock ISSN 0021-9991.
\newblock \doi{http://dx.doi.org/10.1016/j.jcp.2009.06.003}.

\bibitem[Gelder(1998)]{gelder}
Allen~Van Gelder.
\newblock Approximate simulation of elastic membranes by triangulated spring
  meshes.
\newblock \emph{Journal of graphics tools}, 3\penalty0 (2):\penalty0 21--41,
  1998.

\bibitem[Tan et~al.(2012)Tan, Thomas, and Liu]{tan2012}
J.~Tan, A.~Thomas, and Y.~Liu.
\newblock Influence of red blood cells on nanoparticle targeted delivery in
  microcirculation.
\newblock \emph{Soft Matter}, 8\penalty0 (6):\penalty0 1934--1946, 2012.
\newblock \doi{10.1039/c2sm06391c}.

\bibitem[K{\"o}rner et~al.(2006)K{\"o}rner, Pohl, R{\"u}de, Th{\"u}rey, and
  Zeiser]{korner2006}
Carolin K{\"o}rner, Thomas Pohl, Ulrich R{\"u}de, Nils Th{\"u}rey, and Thomas
  Zeiser.
\newblock Parallel lattice {B}oltzmann methods for {CFD} applications.
\newblock In \emph{Numerical Solution of Partial Differential Equations on
  Parallel Computers}, pages 439--466. Springer, 2006.

\bibitem[Schmieschek et~al.(2017)Schmieschek, Shamardin, Frijters, Kr{\"u}ger,
  Schiller, Harting, and Coveney]{lb3d2017}
Sebastian Schmieschek, Lev Shamardin, Stefan Frijters, Timm Kr{\"u}ger, Ulf~D
  Schiller, Jens Harting, and Peter~V Coveney.
\newblock Lb3d: A parallel implementation of the lattice-{B}oltzmann method for
  simulation of interacting amphiphilic fluids.
\newblock \emph{Computer Physics Communications}, 217:\penalty0 149--161, 2017.

\bibitem[Woodgate et~al.(2018)Woodgate, Barakos, Steijl, and
  Pringle]{woodgate2018}
Mark~A Woodgate, George~N Barakos, Rene Steijl, and Gavin~J Pringle.
\newblock Parallel performance for a real time lattice {B}oltzmann code.
\newblock \emph{Computers \& Fluids}, 173:\penalty0 237--258, 2018.

\bibitem[Napolitano and Pascazio(1991)]{pascazio1991}
M~Napolitano and G~Pascazio.
\newblock A numerical method for the vorticity-velocity {N}avier-{S}tokes
  equations in two and three dimensions.
\newblock \emph{Computers \& Fluids}, 19\penalty0 (3-4):\penalty0 489--495,
  1991.

\bibitem[Abraham(1970)]{abraham1970}
Farid~F Abraham.
\newblock Functional dependence of drag coefficient of a sphere on {R}eynolds
  number.
\newblock \emph{The Physics of Fluids}, 13\penalty0 (8):\penalty0 2194--2195,
  1970.

\bibitem[Jeffery(1922)]{jeffery1922}
George~Barker Jeffery.
\newblock The motion of ellipsoidal particles immersed in a viscous fluid.
\newblock \emph{Proceedings of the Royal Society of London. Series A,
  Containing papers of a mathematical and physical character}, 102\penalty0
  (715):\penalty0 161--179, 1922.

\bibitem[Huang et~al.(2012)Huang, Yang, Krafczyk, and Lu]{huang2012}
Haibo Huang, Xin Yang, Manfred Krafczyk, and Xi-Yun Lu.
\newblock Rotation of spheroidal particles in {C}ouette flows.
\newblock \emph{Journal of Fluid Mechanics}, 692:\penalty0 369--394, 2012.

\bibitem[Aidun et~al.(1998)Aidun, Lu, and Ding]{aidun1998}
Cyrus~K Aidun, Yannan Lu, and E-Jiang Ding.
\newblock Direct analysis of particulate suspensions with inertia using the
  discrete {B}oltzmann equation.
\newblock \emph{Journal of Fluid Mechanics}, 373:\penalty0 287--311, 1998.

\bibitem[Zettner and Yoda(2001)]{zettner2001}
CM~Zettner and M~Yoda.
\newblock Moderate-aspect-ratio elliptical cylinders in simple shear with
  inertia.
\newblock \emph{Journal of Fluid Mechanics}, 442:\penalty0 241--266, 2001.

\bibitem[Walter et~al.(2001)Walter, Rehage, and Leonhard]{walter2001}
Anja Walter, Heinz Rehage, and Herbert Leonhard.
\newblock Shear induced deformation of microcapsules: shape oscillations and
  membrane folding.
\newblock \emph{Colloids and Surfaces A: Physicochemical and Engineering
  Aspects}, 183:\penalty0 123--132, 2001.

\bibitem[Li and Sarkar(2008)]{li2008}
Xiaoyi Li and Kausik Sarkar.
\newblock Front tracking simulation of deformation and buckling instability of
  a liquid capsule enclosed by an elastic membrane.
\newblock \emph{Journal of Computational Physics}, 227\penalty0 (10):\penalty0
  4998--5018, 2008.

\bibitem[Knoche and Kierfeld(2011)]{knoche2011}
Sebastian Knoche and Jan Kierfeld.
\newblock Buckling of spherical capsules.
\newblock \emph{Physical Review E}, 84\penalty0 (4):\penalty0 046608, 2011.

\bibitem[Dupont et~al.(2015)Dupont, Salsac, Barthes-Biesel, Vidrascu, and
  Le~Tallec]{dupont2015}
Claire Dupont, A-V Salsac, Dominique Barthes-Biesel, Marina Vidrascu, and
  Patrick Le~Tallec.
\newblock Influence of bending resistance on the dynamics of a spherical
  capsule in shear flow.
\newblock \emph{Physics of Fluids}, 27\penalty0 (5):\penalty0 051902, 2015.

\end{thebibliography}

\end{document}